\RequirePackage{pdf14}

\documentclass[letterpaper]{article}
\pdfoutput=1
\usepackage[utf8]{inputenc}
\usepackage[usenames,dvipsnames,table]{xcolor}
\usepackage{graphicx}
\usepackage{caption}
\usepackage{subcaption}
\usepackage{array,setspace,mathrsfs,amsthm,amsfonts,amsmath,colonequals,mathtools,dsfont}
\usepackage{./auxi/jheppub}
\usepackage{afterpage}
\usepackage{xspace}
\usepackage[normalem]{ulem}
\usepackage{bbm}

\def\x{x}
\def\xb{\bar{\x}}
\def\bzb{{\bf z}}
\def\t{{\theta}}
\def\gap{\mathrm{gap}}
\def\tb{{\bar{\theta}}}
\def\zb{{\bar{z}}}
\def\hb{{\bar{h}}}

\def\ex{\mathrm{ex}}
\def\gsl{g}
\def\sl{sl}

\def\BB{{\mathcal{B}}}
\def\CC{{\mathcal{C}}}
\def\DD{{\mathcal{D}}}
\def\CasEig{\mathfrak{c}}

\def\OO{{\mathcal{O}}}
\def\NN{{\mathcal{N}}}
\def\LL{{\mathcal{L}}}
\def\GG{{\mathcal{G}}}

\def\DD{{\mathcal{D}}}

\def\TT{{\mathcal{T}}}

\def\SU{\mathrm{SU}}
\def\U{\mathrm{U}}

\newcommand{\be}{\begin{equation}}
\newcommand{\ee}{\end{equation}}
\newcommand{\bea}{\begin{eqnarray}}
\newcommand{\eea}{\end{eqnarray}}
\newcommand{\nn}{\nonumber}
\newcommand\qq{\mathbbmtt{Q}}

\newcommand\eg{{\it e.g.}}
\newcommand\ie{{\it i.e.}}

\title{Long Multiplet Bootstrap}
\author{Martina Cornagliotto,}
\author{Madalena Lemos,}
\author{Volker Schomerus}

\affiliation{DESY Hamburg, Theory Group, Notkestra{\ss}e 85, D-22607 Hamburg, Germany}

\emailAdd{$\lbrace$martina.cornagliotto,madalena.lemos,volker.schomerus$\rbrace$@desy.de}

\preprint{DESY 17-026}

\bigskip
\abstract{Applications of the bootstrap program to superconformal field theories
promise unique new insights into their landscape and could even lead to the discovery
of new models. Most existing results of the superconformal bootstrap were obtained form correlation functions of very special fields in short (BPS) representations of the superconformal algebra. Our main goal is to initiate a superconformal bootstrap for
long multiplets, one that exploits all constraints from superprimaries and their
descendants. To this end, we work out the Casimir equations for four-point correlators
of long multiplets of the two-dimensional global $\mathcal{N}=2$ superconformal algebra. After constructing the full set of conformal blocks we discuss two different applications. The first one concerns two-dimensional (2,0) theories. The numerical bootstrap analysis
we perform serves a twofold purpose, as a feasibility study of our long multiplet
bootstrap and also as an exploration of (2,0) theories. A second line of applications is directed towards four-dimensional $\mathcal{N}=3$ SCFTs. In this context, our results imply a new bound $c \geqslant \tfrac{13}{24}$ for the central charge of such models, which we argue cannot be saturated by an interacting SCFT.}


\begin{document}
\pagestyle{myplain}
\setcounter{tocdepth}{2}
\maketitle

\section{Introduction}
\label{sec:intro}

The conformal bootstrap program \cite{Ferrara:1973yt,Polyakov:1974gs,Mack:1975jr} opens a unique new window into the non-perturbative dynamics of quantum field theories, especially since the recent revival through \cite{Rattazzi:2008pe}.
By now we have a large wealth of results on strongly coupled theories, that would otherwise be hard to study
by conventional field theory techniques, even including models that are lacking a Lagrangian description. The
bootstrap approach, by relying only on symmetries, combined with a few spectral assumptions, allows one to obtain
complete non-perturbative answers, without reference to any type of perturbative description. The most striking
example of this is the three-dimensional Ising model, where the best determination of the critical exponents
comes from the bootstrap \cite{ElShowk:2012ht,El-Showk:2014dwa,Kos:2014bka,Simmons-Duffin:2015qma,Kos:2016ysd,
Simmons-Duffin:2016wlq}.
A tremendous amount of work has been done studying superconformal field theories (SCFTs) in various dimensions
and with various amounts of supersymmetry
 \cite{Poland:2010wg,Poland:2011ey,Poland:2015mta,Li:2017ddj,Berkooz:2014yda,
 Beem:2014zpa,Beem:2013qxa,Alday:2013opa,Alday:2014qfa,
 Chester:2014fya,Chester:2014mea,Lemos:2015awa,Chester:2015qca,Bashkirov:2013vya,
 Bobev:2015vsa,Bobev:2015jxa,Beem:2015aoa,Lin:2015wcg,Lin:2016gcl,Lemos:2016xke,
 Dolan:2001tt,Dolan:2004mu,Nirschl:2004pa,Fortin:2011nq,Fitzpatrick:2014oza,Khandker:2014mpa,
 Bissi:2015qoa,Doobary:2015gia,Li:2016chh,Liendo:2015cgi,Liendo:2016ymz}.
It has led to non-perturbative results in known theories ranging from two-dimensional $\NN=(2,2)$ \cite{Lin:2016gcl}, to
 six-dimensional $\NN=(2,0)$ \cite{Beem:2015aoa} SCFTs. By virtue of exploring the space of SCFTs relying only on
symmetries, and with minimum assumptions about the operator content of the theories, the bootstrap program also
provides a way to discover new SCFTs.
Although there have been few surprises so far, a puzzling result was obtained in the supersymmetric bootstrap of
four-dimensional $\NN=1$ SCFT.
Namely the presence of a ``kink'' in the dimension bounds of the leading long operator (\ie, obeying no
shortening conditions) appearing in the operator product expansion (OPE) between a chiral and an antichiral
operator \cite{Poland:2011ey,Poland:2015mta,Li:2017ddj}. Unlike the Ising model case, where the kink appeared exactly
at the location of a known theory, there is no currently known theory which lives at the $\NN=1$ kink.\footnote{
While the presence of a ``kink'' is not enough to guarantee the existence of a fully consistent SCFT, it provides
hints it might correspond to a new $\NN=1$ SCFT. The four-dimensional bounds were extended to SCFTs in $2 \leqslant
d \leqslant 4$ with four-supercharges ($\NN=1$ in four dimensions) \cite{Bobev:2015jxa}, and the ``kink'' persisted in lower dimensions as well. (Although
in fractional dimensions unitarity is not preserved \cite{Hogervorst:2014rta,Hogervorst:2015akt}, the violations
are probably mild as the results appear reasonable.)}
The long operator whose dimension is given by the position of the ``kink'' is one of the natural objects to study
in order to shed light on this ``minimal'' $\NN=1$ SCFT, similarly to what was done for the three-dimensional Ising model.
Very recently the superconformal primary of said long multiplet was considered in \cite{Li:2017ddj}, but the
complete set of constraints arising from the full supermultiplet remains unexplored. The only other existing
bootstrap analysis that went beyond the usual half-BPS multiplets is \cite{Berkooz:2014yda}, but as in
\cite{Li:2017ddj}, the authors restrict to correlations of the superconformal primary.

Most of the study of superconformal field theories (SCFTs) has been limited to the analysis of four-point functions of half-BPS operators.
In this case there are no nilpotent invariants, and the correlation function of the superconformal primary completely
determines that of its superdescendants. Moreover since the only superconformal invariants are the supersymmetrizations
of the conformal and $R$-symmetry cross-ratios, the crossing equations for the superconformal primary four-point function
capture all of the constraints, and there is no need to consider those arising from four-point functions involving
superdescendants. The same is still true for the four-point functions of two chiral operators with two long multiplets
that were studied in \cite{Bobev:2015jxa}. However, things change once we consider four-point functions that involve at
most one half-BPS multiplet while the other fields satisfy fewer or no shortening conditions at all.

\subsubsection*{Long multiplet bootstrap}

For a complete superconformal bootstrap analysis one should certainly consider all four-point functions, including those
in which all fields belong to long multiplets of the superconformal algebra.
Such four-point functions can depend on nilpotent superconformal invariants, and information is lost when restricting the
external operators to the superconformal primaries. For the case of four generic long multiplets this might mean, as was
the case in \cite{Khandker:2014mpa,Li:2016chh,Li:2017ddj} for four-dimensional $\NN=1$ long multiplets, that correlation functions of
superprimaries can (only) be decomposed into bosonic conformal blocks with \textit{independent} coefficients.  While
supersymmetry relates the various operators in the exchanged multiplet, and in particular their conformal dimensions, it
does not constrain the coefficients of the bosonic block decomposition. In other words, correlation functions of
superprimaries in long multiplets possess no ``superblock'' decomposition. The only way the number of free parameters
in these block decompositions may be reduced is through permutation symmetry in the
case of identical fields \cite{Khandker:2014mpa}, or by additional shortening conditions, such as for conserved currents \cite{Fortin:2011nq,Berkooz:2014yda,Khandker:2014mpa,Li:2017ddj}.

In order to fully exploit consequences of supersymmetry in the study of long multiplets, we will be working with the
\emph{full} four-point functions in superspace, \ie, we consider not only superprimaries as the external operators,
but also superdescendants. While our explicit analysis below will focus on two-dimensional SCFTs the key lessons we learn are more
general. We show that, even if there is no ``superblock'' decomposition (other than the one into bosonic blocks) when
one restricts to external superconformal primaries, some of the OPE coefficients of external superdescendants can be
fixed  in terms of those of the primary. This means that the number of free parameters in the block decomposition of the full four-point function is reduced as compared to the decomposition in terms of bosonic blocks. Moreover, the constraints coming from the full
set of crossing equations in superspace are stronger than those of just the superprimary. This is not too surprising
since our approach effectively includes mixed correlators with respect to the bosonic conformal symmetry even if we
analyze correlation functions of four identical supermultiplets of the superconformal algebra. The combination of
a non-trivial superblock decomposition and the constraints from crossing symmetry of superdescendants explains why
our long multiplet bootstrap is significantly more powerful than a conventional analysis of crossing symmetry
for superprimaries in long multiplets.
Recently the aforementioned $\NN=1$ kink was studied by considering simultaneously chiral operators and the \emph{superconformal primary} of long multiplets as external states in the correlation functions \cite{Li:2017ddj}. Even though in this system the blocks corresponding the long four-point function were simply bosonic blocks, stronger results on the kink were obtained. It seems natural to expect an improvement if one adds the (more computationally expensive) \emph{whole} long supermultiplet, and all the crossing symmetry constraints.

In order to illustrate the workings of our long multiplet bootstrap we shall consider models with a
two-dimensional $\NN=2$ (global) superconformal symmetry. Our first goal is to construct the relevant superblocks
for four-point functions of long multiplets. We will do so under some technical assumptions on the
R-charges of the involved multiplets. The superblocks for the various types of exchanged operators,
are obtained in superspace by solving both the quadratic and cubic super Casimir equations. The equations
provided by higher Casimirs bring no new information in this case. We obtain a coupled system of six
second-order differential equations and construct its solutions in terms of hypergeometric functions.
Our analysis serves as a first step towards the computation of long superblocks in higher dimensions
for theories with four supercharges, by solving the super Casimir equation in an arbitrary number of
dimensions, as done in \cite{Bobev:2015jxa} with half-BPS operators. For this reason we focus only on
the global superconformal algebra in two dimensions, and do not make use of the full super
Virasoro algebra.

\subsubsection*{Two-dimensional $\NN=(2,0)$ SCFTs}

Once the relevant superblocks for the $\NN=2$ superconformal algebra are constructed we
can run the numerical bootstrap program for long multiplets. We do so in the context of two-dimensional
$\NN=(2,0)$ SCFTs, putting together the holomorphic blocks we compute with anti-holomorphic global $\sl(2)$
blocks. This serves a two-fold purpose, as a feasibility test of bootstrapping long multiplets, and also as
an exploration of $\NN=(2,0)$ theories which are interesting in their own right.
By focusing on the four-point function of four identical uncharged long multiplets, Bose symmetry fixes all OPE coefficients of
external superdescendants in terms of those of the external superprimary. However the crossing equations for external
superdescendants still provide non-trivial constraints on the CFT data. Indeed if one were to consider the four-point function
of external superconformal primaries alone, one would not find any improvement over the bosonic conformal bootstrap, since there
would be no superblocks as discussed above. We exemplify how the bounds obtained in this way are stronger than the pure bosonic
bootstrap and how our bounds are saturated by known supersymmetric minimal models at a point.

\subsubsection*{Four-dimensional $\NN=3$ SCFTs}

In a different direction, the blocks we have computed are precisely the ones relevant for the study of the chiral
algebras associated to the recently discovered four-dimensional $\NN=3$ SCFTs \cite{Garcia-Etxebarria:2015wns},
further explored in \cite{Aharony:2015oyb,Garcia-Etxebarria:2015wns,Aharony:2016kai,Garcia-Etxebarria:2016erx,
Argyres:2016xua,Argyres:2016xmc,Argyres:2016yzz,Nishinaka:2016hbw,Imamura:2016abe,Imamura:2016udl,Agarwal:2016rvx,Lemos:2016xke}.
Here we take a purely field-theoretic approach to these theories, using the fact, shown in  \cite{Beem:2013sza}, that any four-dimensional theory with
$\NN\geqslant 2$ supersymmetry has a subsector isomorphic to a two-dimensional chiral algebra.
The chiral algebras of $\NN=3$ SCFTs have precisely $\NN=2$ supersymmetry \cite{Nishinaka:2016hbw}.
In the study of four-dimensional four-point functions of half-BPS $\NN=3$ operators, as done in \cite{Lemos:2016xke}, the
relevant two-dimensional blocks are those of external half-BPS (two-dimensional $\NN=2$ chiral) operators, which were computed in
\cite{Fitzpatrick:2014oza}. However, if one wants to consider the four-dimensional stress-tensor multiplet, which
in two dimensions gives rise to the $\NN=2$ stress tensor multiplet, one needs exactly the long blocks obtained
in this paper.
In the spirit of the bootstrap our assumptions will be minimal, obtaining constraints valid for any  \emph{local} and \emph{interacting} $\NN=3$ SCFT.

Therefore we study the four-point function of the stress-tensor multiplet, as it is the only non-trivial multiplet we are guaranteed to have in a local $\NN=3$ SCFT.
We obtain an infinite set of OPE coefficients, between two stress-tensor multiplets and a set of protected operators, valid for any \emph{local, interacting} $\NN=3$ SCFT, depending only on the central charge. This is a necessary first step of any numerical study of the full-blown system of crossing equations for four-dimensional $\NN=3$ stress-tensor multiplets.

Moreover, positivity of these OPE coefficients, as required by unitarity of the four-dimensional $\NN=3$ theory, is not automatic.
Imposing unitarity yields the following analytic bound on the $c$ anomaly coefficient
\be
c_{4d} \geqslant \frac{13}{24}\,,
\ee
valid for any \emph{local, interacting} $\NN=3$ SCFT.
Unlike similar analytic bounds obtained on various central charges, for both $\NN=2$ and $\NN=4$ SCFTs \cite{Beem:2013qxa,Beem:2013sza,Liendo:2015ofa,Lemos:2015orc,Beem:2016wfs}, we argue this bound corresponds to a strict inequality and cannot be saturated by an interacting unitary $\NN=3$ SCFT.

\section{\texorpdfstring{Two-dimensional $\mathcal{N}=2$}{2d N=2} global long superconformal blocks}
\label{sec:superblocks}

In this section we obtain the two-dimensional $\mathcal{N}=2$ global superconformal blocks for the four-point functions of long multiplets. Since the two-dimensional conformal algebra factorizes into left and right movers, we only consider the holomorphic part. Anti-holomorphic blocks will be added in the next section. 
As a warm-up we review the case of the $\NN=1$ blocks \cite{Fitzpatrick:2014oza}, which captures some of the main features of the $\NN=2$ case, while being computationally less involved.
The procedure is as follows: We start by writing the form of the correlation function of four arbitrary operators as required by superconformal symmetry. It will include a general function of the independent superconformal invariants, which amounts to (two) five, for the ($\NN=1$) $\NN=2$ case. The superconformal blocks are then obtained by solving the eigenvalue problem associated to the quadratic and cubic Casimirs.
In the $\NN=2$ case this produces a system of six coupled differential equations for the quadratic Casimir. In order to solve this system we start from a physically motivated 
Ansatz in terms of the expected bosonic block decomposition of the superconformal block.

\subsection{Warm-up example: the \texorpdfstring{$\mathcal{N}=1$}{N=1} superconformal blocks in two dimensions}

We start by revisiting the computation of global superconformal blocks in $\NN=1$ SCFTs \cite{Fitzpatrick:2014oza}, highlighting the main features that are relevant for the $\NN=2$ computation.\footnote{As in \cite{Fitzpatrick:2014oza} we focus on four-point functions that are single-valued on their own, without adding the anti-holomorphic dependence, which reduces the number of invariants and is enough for the illustrative purposes in this section. We thank C.~Behan for pointing this out.}
Recall the global $\NN=1$ superconformal algebra is described, along with the global conformal generators ($L_{\pm1}\,, \; L_0$), by the fermionic generators $G_r$ ($r = \pm 1/2$) with the following commutation relations
\be
\{ G_{r}, G_{s} \}  = 2 L_{r+s}  \ \ \ \mathrm{and } \ \ \ [ L_n, G_{\pm \frac{1}{2}} ] =  \left( \frac{n}{2} \mp \frac{1}{2}  \right) G_{\pm \frac{1}{2} + n } \,.
\ee
Introducing a single fermionic variable $\theta$ we can write a generic superfield $\Phi$, which we label by the holomorphic dimension of its superprimary $h$, as a function of $(x,\theta)$, where $x$ is the usual holomorphic coordinate. The generators can be represented by the differential operators
\be
\begin{alignedat}{2} \label{diff1}
&\LL_{-1} = - \partial_x \,,   \qquad
&&\LL_0 = -x \partial_x - \frac{1}{2} \theta \partial_\theta -h \,, \qquad
\LL_1 = -x^2 \partial_x - x \theta \partial_\theta-2x h \,, \\
&\GG_{-\frac{1}{2}} =  \partial_\theta - \theta \partial_x  \,, \qquad
&&\GG_{+ \frac{1}{2}} = x \partial_\theta - \theta x \partial_x-2 h \theta \,.
\end{alignedat}
\ee
To study the four-point function
\begin{align}
\langle \Phi(x_1, \theta_1)  \Phi(x_2, \theta_2)  \Phi(x_3, \theta_3)  \Phi(x_4, \theta_4)  \rangle \,,
\end{align}
we need to introduce the four-point superconformal invariants on which this correlator can depend. Defining the superconformal distance as $\mathbf{z}_{ij}= x_i - x_j - \theta_i \theta_j$ it is easy to see the two four-point invariants of the theory are
\begin{align}
I_1= \frac{\mathbf{z}_{12} \mathbf{z}_{34}}{\mathbf{z}_{14} \mathbf{z}_{23}} \to \frac{x_1 - x_2 - \theta_1 \theta_2}{x_2} \ \ \ \mathrm{and} \ \ \ \ I_2 =  \frac{\mathbf{z}_{13} \mathbf{z}_{24}}{\mathbf{z}_{14} \mathbf{z}_{23}} \to \frac{x_1}{x_2}\,,
\end{align}
where the arrows mean we used a superconformal transformation to set $\mathbf{z}_3=0 $ and $\mathbf{z}_4 \rightarrow \infty $.
After taking this limit, the four-point function can be written as an arbitrary function of the two invariants
\begin{align}
\label{N=1fourpoint}
G(x_1, x_2, \theta_1, \theta_2)  = \frac{1}{ (x_1 - x_2)^{2h_\phi} }
\left[ g_0 \! \left( z \right) + \frac{\theta_1 \theta_2}{x_2} g_\theta \! \left(z \right) \right] \,,
\end{align}
where $h_\phi$ is the dimension of the superprimary of $\Phi$, $z=1-\frac{x_1}{x_2}$ is the bosonic cross ratio, and of course we used that the Taylor expansion of the function on the fermionic cross ratio truncates.
Let us take a step back to interpret the two functions which appeared: $g_0$ is the piece that survives after taking all fermionic coordinates to zero and thus the four-point function of the superconformal primary of $\Phi$. On the other hand, $g_\theta$ corresponds (up to factors of the bosonic cross ratio) to the correlation function of the two superconformal primaries at points three and four, and two (global) superdescendants at points one and two.

These functions admit a decomposition in blocks, corresponding to the exchange of a given superconformal multiplet.
As in \cite{Fitzpatrick:2014oza}, we obtain these blocks by acting with the 
quadratic Casimir, and solving the corresponding eigenfunction equation, in terms of the eigenvalue of the quadratic Casimir on the exchanged supermultiplet $\CasEig_2$.
The superconformal Casimir $\mathbf{C}^{(d)}$ is given by
\be
\mathbf{C}^{(2)} = L_0^2 - \frac{1}{2} \left( L_1 L_{-1} + L_{-1} L_1 \right ) + \frac{1}{4} \left( G_{+ \frac{1}{2}} G_{-  \frac{1}{2}}   -   G_{-  \frac{1}{2}}  G_{+  \frac{1}{2}} \right)\,.
\ee
Applying the differential form of the 
Casimir on the four-point function \eqref{N=1fourpoint} we obtain a coupled system of two differential equations
\begin{align}
\begin{split}
z^2 \left( (1-z) \partial_z^2 - \partial_z  \right) g_0 + \frac{1}{2} z g_\theta &= \CasEig_2 g_0(z) \,, \\
\left[ z^2  (1-z) \partial_z^2  + z(2-3z) \partial_z  -z + \frac{1}{2} \right] g_\theta + \frac{1}{2} z \left( (1-z) \partial_z^2 - \partial_z  \right) g_0 &= \CasEig_2 g_\theta(z) \,,
\end{split}
\end{align}
where $\CasEig_2 = h(h-\tfrac{1}{2})$ is the eigenvalue of quadratic Casimir on the superconformal multiplet being exchanged, with $h$ denoting the scaling dimension of its superconformal primary.
Solving these equations lead us to two sets of solutions with physical boundary conditions. The first one, obtained in \cite{Fitzpatrick:2014oza}, reads
\begin{align}
\label{solprimary}
\begin{split}
g_0(z) &= \gsl^{0,0}_h \left(\frac{z}{z-1}\right) \,,  \\
g_\theta(z) &= \frac{h}{z} \gsl^{0,0}_h \left(\frac{z}{z-1}\right) \,,
\end{split}
\end{align}
and it corresponds to the case in which the superprimary itself (of weight $h$) is exchanged in the OPE. Note that the argument of the usual $\sl(2)$ block
\be
\label{eq:gsl2}
\gsl^{h_{12},h_{34}}_{h}(z) = z^{h} {}_2 F_1  \left( h - h_{12}, h+h_{34}, 2h, z \right)\,,
\ee
is $\tfrac{z}{z-1}$ since this combination corresponds to the standard bosonic cross-ratio of $\tfrac{x_{12} x_{34}}{x_{13} x_{24}}$.

However, there is a second solution, which has the physical interpretation of a superconformal descendant being exchanged
\begin{align}
\label{soldesc}
\begin{split}
g_0(z) &=\gsl^{0,0}_{h+\tfrac{1}{2}} \left(\frac{z}{z-1}\right)\, ,  \\
g_\theta(z) &= \frac{1-2h}{2z} \gsl^{0,0}_{h+\tfrac{1}{2}} \left(\frac{z}{z-1}\right) \,,
\end{split}
\end{align}
where we recall that $h$ corresponds to the dimension of the superconformal primary, which is what figures in the Casimir eigenvalue.

Notice that if one restricts to the correlation function of superconformal primaries by setting the fermionic coordinates to zero in eq.\ \eqref{N=1fourpoint},  one finds from eqs.\ \eqref{solprimary} and \eqref{soldesc} that the ``superblock'' is a sum of bosonic blocks with arbitrary coefficients. In fact the operator being exchanged in eq.\ \eqref{soldesc} can even be a descendant of an operator which itself does not appear in the OPE decomposition, implying there is not even a constraint on the spectrum. However if one considers the whole supermultiplet as the external field, then one gets superblocks, in the sense that the coefficients of the block decomposition of external superdescendants are fixed in terms of those of external superprimaries.
In practice, by considering the whole superfield as the external operator we are considering a mixed system in which supersymmetry was already used to reduce it to the set of independent of correlators.
Exactly the same will happen for the $\NN=2$ superblocks computed in the remaining of this section.
 As we will see in section \ref{sec:numerics}, for the $\NN=2$ case the crossing equations of external superdescendants provide non-trivial constraints and are essential in obtaining bounds that are stronger than the pure bosonic bootstrap.

\subsection{\texorpdfstring{$\mathcal{N}=2$}{N=2} long multiplet four-point function }
We now apply a similar strategy to the case of $\NN=2$ supersymmetry. Although there are new features with respect to the much simpler $\NN=1$ case, some of the main points are the same, even if obscured by the cumbersome technical details.
The global part of the two-dimensional $\NN=2$ superconformal algebra has four fermionic generators $G_r$, $\overline{G}_r$ ($r=\pm 1/2$),
alongside the standard Virasoro generators $L_m$ ($m=-1,0,1$) and the additional $\U(1)$ R-symmetry current algebra generator $J_0$. The commutation relations are given by
\begingroup
\allowdisplaybreaks[1]

\begin{alignat}{2}
&[L_m, L_n] = (m-n) L_{m+n}\,, \qquad
&&\{ G_r, G_s \} = \{\overline{G}_r ,\overline{G}_s\}  = 0 \,,  \\
&[L_m, J_n] = -n J_{m+n}\,,  \qquad
&&\{ G_r, \overline{G}_s\} = L_{r+s} + \frac{1}{2} (r-s) J_{r+s}\,,   \\
& [L_m,G_r ] = \left( \frac{m}{2} - r \right) G_{r+m} \,, \qquad
&&[J_m, G_r] =G_{m+r} \,, \\
& [L_m,\overline{G}_r ] = \left( \frac{m}{2} - r \right) \overline{G}_{r+m} \,, \qquad
&& [J_m, \overline{G}_r] = - \overline{G}_{m+r} \,.
\end{alignat}
\endgroup
Introducing  $\theta$ and $\bar{\theta}$  as the two fermionic directions, following the steps of the previous subsection, we start by writing  the differential action of the generators as (see for example \cite{Kiritsis:1987np})
\begingroup
\allowdisplaybreaks[1]
\begin{align}
\label{difform}
\begin{split}
&\mathcal{L}_{-1} = -\partial_x\,, \\
&\mathcal{L}_0 =- x \partial_x - \frac{1}{2} \t \partial_{\t} - \frac{1}{2} \tb \partial_{\tb} -h\,, \\
&\mathcal{L}_1 = -x^2 \partial_x - x \t \partial_{\t} - x \tb \partial_{\tb} -2xh + q\t \tb \,, \\
&\mathcal{G}_{+\tfrac{1}{2}} = \frac{1}{\sqrt{2}} \big( x \partial_{\t} + \t \tb \partial_{\t} -x \tb  \partial_x- (2h +q) \tb \big)\,, \\
&\overline{\mathcal{G}}_{+\tfrac{1}{2}} = \frac{1}{\sqrt{2}}\big( x \partial_{\tb} - \t \tb \partial_{\tb} -x  \t  \partial_x- (2h-q)\t \big) \,, \\
&\mathcal{G}_{-\tfrac{1}{2}} = \frac{1}{\sqrt{2}} (\partial_{\t} -  \tb \partial_x )\,, \\
&\overline{\mathcal{G}}_{-\tfrac{1}{2}} = \frac{1}{\sqrt{2}}  (  \partial_{\tb}-\tb \partial_x ) \,, \\
&\mathcal{J}_0 =- \t \partial_{\t} + \tb \partial_{\tb} - q\,,
\end{split}
\end{align}
\endgroup
where $h$ and $q$ are the conformal weight and the R-charge respectively, of the superconformal primary of the superfield.

\subsubsection*{Superconformal invariants}

The form of the long multiplet four-point function is fixed by superconformal invariance up to an arbitrary function of all four-point superconformal invariants.  Defining the supersymmetric distance
\be
Z_{ij}=x_i-x_j-\t_i \tb_j-\tb_i \t_j\,, \quad \text{with } \t_{ij}= \t_i  - \t_j\,,
\ee
there are five such invariants, most naturally written as\footnote{Such invariants also appeared in \cite{Kiritsis:1987np} although only four were considered independent there. However, as will become clear later, all five invariants we write are independent and required for the four-point function expansion.}
\begin{align}
\label{eq:Uinv}
\begin{alignedat}{2}
&U_1= \frac{Z_{13}Z_{24}}{Z_{23}Z_{14}} \,,
&&U_4= \frac{\t_{12}\tb_{12}}{Z_{12}}+\frac{\t_{24}\tb_{24}}{Z_{24}}-\frac{\t_{14}\tb_{14}}{Z_{14}} \,, \\
&U_2= \frac{\t_{13}\tb_{13}}{Z_{13}}+\frac{\t_{34}\tb_{34}}{Z_{34}}-\frac{\t_{14}\tb_{14}}{Z_{14}} \,,\qquad
&&U_5= \frac{Z_{12}Z_{34}}{Z_{23}Z_{14}} \,, \\
&U_3= \frac{\t_{23}\tb_{23}}{Z_{23}}+\frac{\t_{34}\tb_{34}}{Z_{34}}-\frac{\t_{24}\tb_{24}}{Z_{24}}  \,.
\end{alignedat}
\end{align}
With foresight we define a new basis of invariants $I_a$ as
\begin{align}
\label{eq:Iinv}
\begin{alignedat}{3}
&I_0= 1-U_1 \,, \qquad &&I_1=-U_5-(1-U_1)\,,  \qquad  &&   \\
&I_2= U_4(1-U_1)+U_2 U_1 \,,\qquad &&I_3=U_3 \,,  \qquad && I_4=U_2\,. 
\end{alignedat}
\end{align}
Naturally, all these invariants should be nilpotent at some power, with the exception of the one that corresponds to the supersymmetrization of the bosonic conformal invariant. Indeed we find they obey the following identities
\begin{align}
\label{eq:Invprops}
\begin{alignedat}{3}
& I_1^2=-2 I_3 I_4 (1-I_0)\,,\ \ \ \ \ \ \ \ &&I_2^2= 2 I_3 I_4 (1-I_0)\,, \ \ \ \ \ \ \ \ &&I_3^2=0\,,  \\
& I_1 I_2 =I_1 I_3 =I_1 I_4=0\,,\ \ \ \ \ \ \ \ && I_2 I_3 = I_2 I_4=0\,, \ \ \ \ \ \ \  && I_4^2=0\,,
\end{alignedat}
\end{align}
with the non-nilpotent invariant being $I_0$. The four-point function then has a finite Taylor expansion in the nilpotent invariants, with each term being a function of the super-symmetrization of the bosonic cross ratio $I_0$.

\subsubsection*{The four-point function}

We write a generic long  $\NN=2 $ superconformal multiplet as
\be
\label{superfield}
 \Phi(x, \t,\tb) =  \phi(x) + \t \psi(x) + \tb \chi(x) + \t \tb T(x)\,,
\ee
and we label the multiplet by the quantum numbers of its superconformal primary $\phi(x)$, namely the R-charge, $q$, and holomorphic dimension $h$. In our conventions then $\psi(x)$ ($\chi(x)$) has dimension and charge  $h+\tfrac{1}{2}$ and $q+1$ ($h+\tfrac{1}{2}$ and $q-1$), while $T(x)$ has charge $q$ and dimension $h+1$. Notice also that $T(x)$  is, in general, not a conformal primary, since it is not annihilated by the special conformal transformations. The superdescendant of dimension $h+1$ that is a conformal primary  corresponds to the combination $P=-T- \tfrac{q}{2h} \partial \phi(x)$.\footnote{Note that since we are working only with the global part of the conformal algebra, by conformal primary we do not mean a Virasoro primary but rather what is sometimes called a quasi-primary.}

We now write the most general form of the four-point function, as required by superconformal invariance.
As usual we write the correlation function as a prefactor
carrying the appropriate conformal weights, times a function of superconformal invariants
\be
\langle \Phi(x_1, \t_1,\tb_1) \Phi(x_2, \t_2,\tb_2) \Phi(x_3, \t_3,\tb_3)\Phi(x_4, \t_4,\tb_4)\rangle =
\frac{1+ \frac{q_1 \t_{12} \tb_{12}}{Z_{12}}}{Z_{12}^{2h}}\frac{1+ \frac{q_3 \t_{34} \tb_{34}}{Z_{34}}}{Z_{34}^{2h'}}F(I_a) \,,
\label{eq:fourpt}
\ee
where for simplicity we took $h_1=h_2=h$, $h_3=h_4=h'$, $q_2=-q_1$ and $q_4=-q_3$ for the conformal dimensions and charges of the superprimaries.
Given the properties of the nilpotent invariants \eqref{eq:Invprops} the function $F(I_a)$ can be expanded as
\begin{align}
\label{eq:susyexp}
F(I_a) = f_0(I_0)+ I_1 f_1(I_0)+I_2 f_2(I_0)+I_3f_3(I_0)+I_4(1-I_0)f_4(I_0) + I_3 I_4 (1-I_0)f_5(I_0)\,.
\end{align}
Furthermore we can use a superconformal transformation to set $x_4= \infty$, $x_3=0$, and the fermionic variables of the last two fields to zero yielding
\begin{align}
\begin{split}
\label{Fexp}
F(z,\t_\alpha,\tb_\alpha) =& f_0(z)+\frac{f_1(z) (\theta _1\bar{\theta}_2-\theta _2\bar{\theta}_1)}{x_2}+\frac{f_2(z) (\theta _1\bar{\theta}_2+\theta _2\bar{\theta}_1)}{x_2}+\frac{f_3(z) \theta _2\bar{\theta}_2}{x_2}+\frac{f_4(z) \theta _1\bar{\theta}_1}{x_2} \\
&-\frac{f_5(z) \theta _1 \theta _2\bar{\theta}_1\bar{\theta}_2}{x_2^2}\,,
\end{split}
\end{align}
with  $\alpha=1,2$ and $z=1-\frac{x_1}{x_2}$.\footnote{\label{zfoot} Notice again that the standard two-dimensional cross-ratio is related to $z$ by $\bzb=\tfrac{z}{z-1}=\tfrac{x_{12} x_{34}}{x_{13} x_{24}}$.}
The natural form of equation \eqref{Fexp} is what motivated the choice of invariants $I_a$ in eq.\ \eqref{eq:Iinv}.
Our goal is to obtain how each superconformal multiplet appearing in the double OPE of $\Phi$ contributes to each of these functions $f_i(z)$.
Following the warm-up example the next step is to write the 
Casimir operators and act with them on the correlation function.

\subsubsection*{Quadratic and cubic Casimirs}

In the case under investigation it turns out that the quadratic Casimir $\mathbf{C}^{(2)}$ is not enough to completely fix the form of the superconformal blocks, and we must also use the cubic one $\mathbf{C}^{(3)}$.
This can readily be seen by looking at the eigenvalue of the quadratic Casimir on a superconformal multiplet whose superprimary has dimension $h_{\ex}$ and charge $q_{\ex}$,
\be
\CasEig_2=h_{\ex}^2 - \tfrac{q_{\ex}^2}{4}\,,
\label{eq:QuadrCasEig}
\ee
which does not distinguish the sign of the R-charge.
Superconformal multiplets with opposite charges are distinguished by the cubic Casimir, whose eigenvalue is
\be
\CasEig_3=-q_{\ex} \CasEig_2\,.
\ee
The quadratic and cubic Casimirs are given by \cite{Arnaudon:1995mf}
\begin{align}
\label{eq:Cas2}
\mathbf{C}^{(2)} &= L_0^2 - \frac{1}{4} J_0^2 - \frac{1}{2} \{ L_1, L_{-1} \} + \frac{1}{2} [\overline{G}_{+\tfrac{1}{2}}, G_{-\tfrac{1}{2}}]+ \frac{1}{2}[G_{+\tfrac{1}{2}},\overline{G}_{-\tfrac{1}{2}}] \,,\\
\label{Cas3}
\begin{split}
\mathbf{C}^{(3)} = \ & (L_0^2 - \frac{1}{4} J_0^2 - \frac{1}{2}  L_{-1} L_1) J_0 + G_{-\tfrac{1}{2}}\overline{G}_{+\tfrac{1}{2}}(1-L_0-\frac{3}{2} J_0)  \\
&-\overline{G}_{-\tfrac{1}{2}}G_{+\tfrac{1}{2}}(1-L_0+\frac{3}{2} J_0) -L_{-1} \overline{G}_{+\tfrac{1}{2}}G_{+\tfrac{1}{2}} +L_{1} G_{-\tfrac{1}{2}}\overline{G}_{-\tfrac{1}{2}}\, .
\end{split}
\end{align}
Acting with the quadratic Casimir on the four point function, through the differential action \eqref{difform} of the generators, yields a system of six coupled differential equations for the six functions $f_i(z)$ in eq.\ \eqref{Fexp}.
These are rather long and thus we collect them in appendix \ref{app:quadraticCasimireqs}; the next step is find a solution for this system of coupled differential equations, and then to constrain said solution further by demanding it is also an eigenfunction of the cubic Casimir equation.

\subsection{Long superconformal blocks }
\label{sec:longblockssol}

The easiest, and physically more transparent, way to solve the system of Casimir equations given in appendix \ref{app:quadraticCasimireqs} is to give an Ansatz in terms of the expected bosonic block decomposition of superblocks.\footnote{Note that by giving an Ansatz as a sum of bosonic blocks we are already fixing the boundary conditions and don't have to worry about removing shadow-block solutions.}
Instead of given an Ansatz for the functions $f_i$, it is more convenient  to ``change basis'' from the $f_i$ to functions  $\hat{f}_i$
that match the individual four-point functions of each external superconformal descendant (but conformal primary) field. This change of basis reads
\begingroup
\allowdisplaybreaks[1]
\begin{align}
&f_ 0 (z) = \hat {f_ 0} (z) \,,\nn \\
&f_ 2 (z) = -\frac {2 q_ 1\hat{f_ 0} (z) + \hat{f_ 1} (z) + \hat {f_ 2} (z)} {2 z} \,,\nn \\
&f_ 1 (z) = \frac {4h\hat {f_ 0} (z) + \hat {f_ 1} (z) - \hat {f_ 2} (z)} {2 z} \,,\nn \\
\label{eq:ftofhat}
&f_ 3 (z) = \frac {2h\hat {f_ 3} (z) - q_ 1 (z - 1) z\hat {f_ 0}' (z)} {2h z} \,, \\
&f_ 4 (z) = \frac {q_ 1\hat {f_ 0}' (z)} {2h} + \frac {\hat {f_ 4} (z)} {z} \,,\nn \\
&f_ 5 (z) = \frac{1}{4h^2 z^2} \Big( 2h (2h - 1)\left (4h^2 -  q_1^2 \right)\hat {f_ 0} (z) -q_ 1^2 z^2(\hat {f_ 0}' (z)-(1-z) \hat {f_ 0}'' (z))-\nn \\
& - 4h^2 \big((q_1 - 2h)\hat{f_ 1} (z)- (q_1+2h)\hat{f_ 2} (z)-\hat{f_ 5}  \big)+2h q_ 1 z(\hat{f_ 3}(z)-\hat{f_ 3}' (z)) +2h q_ 1 z(1-z)\hat{f_ 4}' (z) \Big)\,.\nn
\end{align}
\endgroup
This was obtained by expanding the superfields \eqref{superfield} on the left-hand-side of the four-point function \eqref{eq:fourpt}, and obtaining the combinations of $f_i$ that captures the correlation function of each of the conformal primaries appearing in eq.\ \eqref{superfield}.\footnote{Note that $T$ in eq.\ \eqref{superfield} is not the conformal primary combination as discussed below that equation.}

Each of the functions $\hat{f}_i$ then has the interpretation as corresponding the correlators listed in eq.\ \eqref{eq:hatf_ans}, and admits a decomposition in regular bosonic blocks. Recall that the Casimir equations depend on the quantum numbers of the superprimary of the multiplet being exchanged: $q_{\ex}$ and $h_{\ex}$. As such the most generic contribution of a given multiplet may be decomposed into a sum of bosonic blocks
with dimensions that are determined by the dimensions of the various fields in 
the multiplet,
\begingroup
\allowdisplaybreaks[1]
\begin{align}
\label{eq:hatf_ans}
\begin{split}
&\left.\hat{f_0} \right|_{h_{\ex}} = a_0 \gsl^{0,0}_{h_{\ex}} +b_0 \gsl^{0,0}_{h_{\ex}+1} +c_0 \gsl^{0,0}_{h_{\ex}+\tfrac{1}{2}} \rightarrow \langle \phi_1 \phi_2 \phi_3 \phi_4 \rangle \,,  \\
&\left.\hat{f_1} \right|_{h_{\ex}} = a_1 \gsl^{0,0}_{h_{\ex}} +b_1 \gsl^{0,0}_{h_{\ex}+1} +c_1 \gsl^{0,0}_{h_{\ex}+\tfrac{1}{2}}  \rightarrow -\langle  \chi_1 \psi_2 \phi_3 \phi_4 \rangle \,,  \\
&\left.\hat{f_2} \right|_{h_{\ex}}  = a_2 \gsl^{0,0}_{h_{\ex}} +b_2 \gsl^{0,0}_{h_{\ex}+1} +c_2 \gsl^{0,0}_{h_{\ex}+\tfrac{1}{2}} \rightarrow \langle \psi_1 \chi_2 \phi_3 \phi_4 \rangle \,, \\
&\left.\hat{f_3}  \right|_{h_{\ex}} = a_3 \gsl^{-1,0}_{h_{\ex}} +b_3 \gsl^{-1,0}_{h_{\ex}+1} +c_3 \gsl^{-1,0}_{h_{\ex}+\tfrac{1}{2}} \rightarrow \langle \phi_1 P_2 \phi_3 \phi_4 \rangle \,, \\
&\left.\hat{f_4}  \right|_{h_{\ex}} = a_4 \gsl^{1,0}_{h_{\ex}} +b_4 \gsl^{1,0}_{h_{\ex}+1} +c_4 \gsl^{1,0}_{h_{\ex}+\tfrac{1}{2}} \rightarrow \langle P_1 \phi_2 \phi_3 \phi_4 \rangle \,,  \\
&\left.\hat{f_5} \right|_{h_{\ex}}  = a_5 \gsl^{0,0}_{h_{\ex}} +b_5 \gsl^{0,0}_{h_{\ex}+1} +c_5 \gsl^{0,0}_{h_{\ex}+\tfrac{1}{2}} \rightarrow  \langle P_1 P_2 \phi_3 \phi_4 \rangle \,.
\end{split}
\end{align}
\endgroup
Where again $\gsl^{h_{12},h_{34}}_{h_{\ex}}$ is the standard $\sl(2)$ conformal block, \eqref{eq:gsl2}, with argument $\frac{z}{z-1}$ (see footnote \ref{zfoot}).
Next we note that since we are considering the OPE channel between two oppositely charged fields, by $\U(1)$ R-charge conservation only uncharged operators can appear.
Then we have two possibilities
\begin{itemize}
\item The superconformal primary itself is uncharged ($q_{\ex}=0$), which means both the primary (the exchange with coefficient $a_i$ in eq.\ \eqref{eq:hatf_ans}) and its dimension $h_{\ex}+1$ superdescendant can appear ($b_i$ in eq.\ \eqref{eq:hatf_ans}), but not its dimension $h_{\ex + \tfrac{1}{2}}$ and thus $c_i=0$,
\item The superconformal primary has charge $q_{\ex}=\pm 1$, which means only one of its two dimension $h_{\ex}+\tfrac{1}{2}$ can appear (the exchange with coefficient $c_i$ in eq.\ \eqref{eq:hatf_ans}), and thus $a_i=b_i=0$.
\end{itemize}
This is in accord with the study of $\NN=2$ three-point functions of \cite{Blumenhagen:1992sa}. For the exchange of a given supermultiplet labeled by $q_{\ex}$ and $h_{\ex}$ the various coefficients in eq.\ \eqref{eq:hatf_ans} are constrained by the Casimir equations.

\subsubsection*{Uncharged supermultiplet exchange}

First we consider the solutions where the superconformal primary has zero charge, in which case $c_i=0$ in eq. \eqref{eq:hatf_ans}.
Plugging the Ansatz \eqref{eq:hatf_ans} in the quadratic Casimir equations for $\hat{f}_i$, obtained from the ones in appendix \ref{app:quadraticCasimireqs}, we find the following solution
\begingroup
\allowdisplaybreaks[1]
\begin{align}
\label{eq:atypesol}
\begin{split}
a_2&= 4 a_0 \left(h -\frac{h_{\ex}}{2}\right)+a_1\,,  \\
a_3&= a_1-\frac{a_0 \left(-2 h ^2+h  h_{\ex}-h  q_1+\frac{1}{2} h_{\ex} q_1\right)}{h }\,,  \\
a_4&= a_1-\frac{a_0 \left(-2 h ^2+h  h_{\ex}-h  q_1+\frac{1}{2} h_{\ex} q_1\right)}{h }\,,  \\
a_5&= \frac{a_0 \left(2 h ^2-h  (2 h_{\ex}-1)+\frac{1}{2} (h_{\ex}-1) h_{\ex}\right) (2 h +q_1)^2}{2 h ^2}+\frac{a_1 q_1 (2 h -(h_{\ex}-1))}{h }\,,
\end{split}
\\
\label{eq:btypesol}
\begin{split}
b_2&= 4 b_0 \left(h +\frac{h_{\ex}}{2}\right)+b_1\,,  \\
b_3&= -\frac{b_1 (h_{\ex}+1)}{h_{\ex}}-\frac{b_0 (h_{\ex}+1) \left(h +\frac{h_{\ex}}{2}\right) (2 h +q_1)}{h  h_{\ex}}\,,  \\
b_4&= -\frac{b_1 (h_{\ex}+1)}{h_{\ex}}-\frac{b_0 (h_{\ex}+1) \left(h +\frac{h_{\ex}}{2}\right) (2 h +q_1)}{h  h_{\ex}}\,,  \\
b_5&= \frac{b_0 \left(2 h ^2+h  (2 h_{\ex}+1)+\frac{1}{2} h_{\ex} (h_{\ex}+1)\right) (2 h +q_1)^2}{2 h ^2}+\frac{b_1 q_1 (2 h +h_{\ex}+1)}{h }\,,
\end{split}
\end{align}
\endgroup
where one of the unfixed $a_i$, and one of the unfixed $b_i$ correspond to normalizations, thus leaving one arbitrary parameter in each of the above solutions.
This solution automatically solves the constraints coming from the cubic and quartic Casimirs, and thus no more parameters can be fixed in general. Until now we were considering arbitrary fields of pairwise equal charges, however the parameters can be further constrained if the operators are assumed to be conjugates of each other.
If, in addition, we consider the case $\Phi_1 = \Phi_2$, $\Phi_3 = \Phi_4$, for which we need to consider uncharged operators ($q_1=-q_2=q_3=-q_4=0$), we notice that, for example, $a_1$ and $a_2$ correspond  to the same three-point functions up to a permutation of the first two fields. Therefore imposing Bose symmetry fixes two more parameters as
\be
a_1=- a_0 (2 h -h_{\ex})\,, \qquad \text{and } \qquad b_1= - b_0 (2 h +h_{\ex})\,.
\label{eq:identicalops}
\ee
In this case then, the contribution of a given operator to the OPE of the descendants of the external superfield is fixed in terms of that of the external primary operator.

\subsubsection*{Charged supermultiplet exchange}

Finally, we turn to the exchange of a multiplet whose superconformal primary has charge $q_{\ex}=\pm 1$.
By solving the quadratic Casimir equation we clearly cannot distinguish the sign of $q_{\ex}$ and thus we must also consider the cubic Casimir.
Unlike the quadratic case, the eigenvalue  $\CasEig_3=-q_{\ex} \CasEig_2$ depends on the sign of $q_{\ex}$,  and this allows to fix completely all of the $c_i$, in terms of the charge of the exchanged supermultiplet $q_{\ex}=\pm 1$, to be
\begingroup
\allowdisplaybreaks[1]
\begin{align}
c_1&= -c_0 (2 h +q_1)\,, \nn \\
c_2&= -c_0 (q_1-2 h )\,, \nn \\
c_3&= \frac{-q_{\ex} (c_0 (2 h_{\ex}+1) (2 h +q_1 q_{\ex}))}{4 h }\,, \nn \\
\label{eq:ctypesol}
c_4&= -\frac{-q_{\ex} (c_0 (2 h_{\ex}+1) (2 h -q_1 q_{\ex}))}{4 h }\,, \\
c_3&= \frac{c_0 (2 h_{\ex}+1) (2 h -q_1)}{4 h }\,, \nn \\
c_5&= -\frac{c_0 \left(-64 h ^4-32 h ^3+4 h ^2 \left(4 h_{\ex}^2-1\right)+16 h ^2 q_1^2+8 h  q_1^2-\left(4 h_{\ex}^2-1\right) q_1^2\right)}{16 h ^2}\,,\nn
\end{align}
\endgroup
where $c_0$ is a normalization.
This solution automatically satisfies the quartic Casimir equation. The final system of Casimir equations are collected in eq.\ \eqref{quarticeq}.

\subsection{Decomposition of the \texorpdfstring{$\NN=2$}{N=2} stress-tensor four-point function}
\label{sec:STdecomp}

In order to prepare for our analysis of four-dimensional $\NN=3$ theories in section 
\ref{sec:Neq3}, and as a consistency check for the superblocks we constructed, we want 
to decompose the $\NN=2$ stress-tensor four-point function in terms of our blocks \eqref{eq:hatf_ans}.
The stress-tensor multiplet of an $\NN=2$ superconformal field theory in two dimensions is composed of the  $\U(1)$ current, $J(x)$,  two fermionic supercurrents, $G(x)$ and $\bar{G}(x)$, and the stress tensor itself $T(x)$. These four currents can be naturally organized in a long supermultiplet
\be
\mathcal{T}(x,\theta,\bar{\theta})= J(x) + \t G(x) + \tb \bar{G}(x) + \t \tb T(x)\,,
\ee
whose superprimary has dimension one and charge zero. Therefore, the four point function
\be
\langle \mathcal{T}(x_1,\t_1,\tb_1) \mathcal{T}(x_2,\t_2,\tb_2)\mathcal{T}(x_3,\t_3,\tb_3)\mathcal{T} (x_4,\t_4,\tb_4)\rangle\,,
\ee
corresponds precisely to the type we have studied in this section, and it admits a decomposition in the blocks we just computed.
Thus, after fixing the following list of correlators, for example by using Ward identities, we can decompose them in blocks of eq.\ \eqref{eq:hatf_ans} as\footnote{The minus signs may appear strange but they just follow from the way $\hat{f}_i$ was defined, and so are to be combined with the corresponding $a_i$, $b_i$ and $c_i$ in eq.\ \eqref{eq:hatf_ans}.}
\begin{align}
\begin{split}
\langle J(x_1)J(x_2)J(x_3)J(x_4) \rangle = \sum_{h_{\ex}} \lambda^2_{ex} \hat{f}_0(h_{\ex}) \,,\\
\langle \bar{G}(x_1) G(x_2) J(x_3) J(x_4) \rangle = -\sum_{h_{\ex}} \lambda^2_{ex} \hat{f}_1(h_{\ex}) \,, \\
\langle G(x_1) \bar{G}(x_2) J(x_3)J(x_4) \rangle = \sum_{h_{\ex}} \lambda^2_{ex} \hat{f}_2(h_{\ex}) \,,\\
\langle T(x_1) T(x_2) J(x_3)J(x_4) \rangle =\sum_{h_{\ex}} \lambda^2_{ex} \hat{f}_5(h_{\ex}) \,, \\
\end{split}
\end{align}
while the contributions of $\hat{f}_3$ and $\hat{f}_4$ have to amount to zero, since the corresponding correlator vanishes.
Note that we assumed the four $J(x)$ currents to be equal, and thus Bose symmetry requires the $\sl(2)$ block decomposition of the four identical currents to be in terms of exchanged operators with even holomorphic dimension. This constraints the dimension of the superprimary of the exchanged operators, $h_{\ex}$, for the three types of solutions, as can be read from eq.\ \eqref{eq:hatf_ans}. In turn, this implies that for uncharged exchanges either the superconformal primary ($a$ in eq.\ \eqref{eq:hatf_ans}) or its descendent ($b$ in eq.\ \eqref{eq:hatf_ans}) appear, but not both at the same time. Moreover, Bose symmetry fixes the unfixed coefficients in eqs.\ \eqref{eq:atypesol} and \eqref{eq:btypesol} according to eq.\ \eqref{eq:identicalops}. Therefore we label the OPE coefficient $\lambda_{ex}(X)$ with $X=a,b,c$ according to which of the solutions in eqs.\ \eqref{eq:atypesol}, \eqref{eq:btypesol} and \eqref{eq:ctypesol} is turned on, and for the corresponding one we normalize $X_0=1$.
Doing so we find
\begingroup
\allowdisplaybreaks[1]
\begin{align}
\label{eq:STdecinsuper}
\begin{split}
&\lambda^2_{\ex}(a)= 1 \,, \; h_{\ex}=0 \,,\\
&\lambda^2_{\ex}(a)= \frac{\sqrt{\pi } 2^{1-2  h_{\ex}} \left( c (h_{\ex}-1) \Gamma (h_{\ex}+3)+12 ((h_{\ex}-2) h_{\ex}-2) \Gamma (h_{\ex}) \right)}{c  h_{\ex} \Gamma
   \left( h_{\ex}+\frac{1}{2}\right)}\,, \; h_{\ex}=2k\,,\\
&\lambda^2_{\ex}(b)=
\frac{\sqrt{\pi } 2^{-2  h_{\ex}-1} \left( c (h_{\ex}-2) (h_{\ex}-1) h_{\ex} (h_{\ex}+1)+12 (h_{\ex} (h_{\ex}+2)-2)\right) \Gamma
   \left( h_{\ex}\right)}{c  h_{\ex} \Gamma
   \left( h_{\ex}+\frac{3}{2}\right)}\,, \; h_{\ex} =2k-1\,,\\
&\lambda^2_{\ex}(c, q_{\ex}=1)=
\frac{\sqrt{\pi } 2^{-2  h_{\ex}-3} \left(4  h_{\ex}^2-9\right)
   \left(4 c  h_{\ex}^2-c-48\right) \Gamma
   \left( h_{\ex}-\frac{1}{2}\right)}{c \left(2  h_{\ex}+1\right)
   \Gamma \left( h_{\ex}\right)} \,, \; h_{\ex} = \frac{4k-1}{2} \,,\\
&\lambda^2_{\ex}(c, q_{\ex}=-1)=
\frac{\sqrt{\pi } 2^{-2  h_{\ex}-3} \left(4  h_{\ex}^2-9\right)
   \left(4 c  h_{\ex}^2-c-48\right) \Gamma
   \left( h_{\ex}-\frac{1}{2}\right)}{c \left(2  h_{\ex}+1\right)
   \Gamma \left( h_{\ex}\right)}\,, \; h_{\ex} = \frac{4k-1}{2} \,.
\end{split}
\end{align}
\endgroup
where the first term corresponds to the identity contribution, and below $k$ is a positive integer.
All these OPE coefficients are positive for unitarity theories, as they correspond, in our normalizations, to OPE coefficient squared between two identical currents and a generic operator. For negative central charges, \ie, non-unitary theories, of course this is no longer the case and this will be crucial to constrain the space of four-dimensional SCFTs following \cite{Beem:2013sza} in section~\ref{sec:Neq3}.

\section{\texorpdfstring{Bootstrapping two-dimensional $(2,0)$ theories}{Bootstrapping two-dimensional (2,0) theories}}
\label{sec:numerics}

As a first application of our bootstrap program for long operators we shall consider two-dimensional theories
with $\NN=(2,0)$ supersymmetry. The superblocks we constructed in the previous section suffice to analyze
constraints from crossing symmetry of uncharged fields. In addition, we shall assume that our four external
fields are identical and scalar ($h=\hb$). These two assumptions would be easy to drop, but they simplify
things a bit. In particular, for \emph{identical} uncharged scalar operators the contribution from any given
supermultiplet is determined by a single OPE coefficient, as shown in section \ref{sec:longblockssol}.

Below we shall briefly review the history and status of $\NN=(2,0)$ theories before we work out the
crossing symmetry constraints. We then combine the decomposition into our left moving superblocks
with a standard decomposition into right moving bosonic blocks to prepare for a numerical analysis.
The results on bounds for central charges and conformal weights are summarized in the final subsection.
Let us note that any $\NN=(2,2)$ theory is an $\NN=(2,0)$ one so that we cannot remove these solutions to
the crossing equations from our analysis. In particular, we will see how we can recover models with
more supersymmetry within the smaller system of crossing equations of $\NN=(2,0)$ theories. One
clear example is the $k=2$ minimal model with $\NN=(2,2)$. While this model seemed to appear \emph{inside}
the region that is allowed by crossing symmetry of chiral operators, for the truncation of this system considered in \cite{Bobev:2015vsa}, the
central charge bounds in our long multiplet bootstrap are such that the $k=2$ minimal model actually
saturates them.

\subsection{\texorpdfstring{The landscape of two-dimensional $\NN=(2,0)$ theories}{The landscape of two-dimensional N=(2,0) theories}}
\label{sec:landscape}

The study of two-dimensional models with (2,0) supersymmetry goes back more than
two decades. Originally the main motivation came from heterotic string theory which
relies on worldsheet models in which left movers are acted upon by an $\mathcal{N}
=2$ superconformal algebra while right movers carry an action of the Virasoro algebra
only. Extending the simplest realizations of this setup, which involves free fields, to
non-trivial curved backgrounds turned out much more difficult than in the case of
(2,2) supersymmetry. This has two reasons. On the one  hand, the reduced amount
of supersymmetry provides less control over the infrared  fixed points of renormalization
group flows in potential two-dimensional gauge theory realizations.  On the other hand, exact
worldsheet constructions need to adapt to the fact that  left- and right movers are
not identical, an issue that could be overcome in a few cases which we will describe
below. But even with some exactly solvable models around, it remains an open
question how typical they are within the landscape of two-dimensional (2,0) theories.
More recent developments provided a new view onto this landscape. In fact, a large
family of (2,0) theories are expected to emerge when one wraps $M5$ branes on a
4-manifold $\mathcal{M}_4$ \cite{Gadde:2013sca}. Thereby the rich geometry of
4-manifolds becomes part of the landscape of two-dimensional (2,0) theories.

Realization as infrared fixed points of two-dimensional gauge theories were initiated by
Witten in \cite{Witten:1993yc}. In this paper, some gauged linear sigma models for
(2,2) theories are deformed by terms that break the right moving supersymmetry. These
could be shown to flow to conformal field theories in the infrared \cite{Distler:1994hq}.
The framework was extended to a larger class of gauged linear sigma models in
\cite{Distler:1993mk} and arguments for the existence of infrared fixed points
were given in \cite{Silverstein:1995re}. More recently, realizations of (2,0) theories
that are based on two-dimensional \textit{non-abelian} gauge theories were pioneered in
work of Gadde et al. \cite{Gadde:2013lxa}.  Within this extended setup, interesting new
non-perturbative triality relations emerged. Controlling the infrared behavior of these
theories, however, remains a tricky issue, even with the use of modern technology
\cite{Gadde:2014ppa}.

Soon after the early work in the context of gauged linear sigma models, the first
families of exact solutions were constructed in \cite{Berglund:1995dv}, following
earlier ideas in \cite{Giddings:1993wn}, see also  \cite{Gannon:1992np,Gannon:1994vp,
Blumenhagen:1995tt}. In all existing constructions, the left moving $\mathcal{N}=2$
sector is realized as a gauged (coset) WZNW model, following the work of Kazama and
Suzuki \cite{Kazama:1988qp}. The simplest realization was found in
\cite{Berglund:1995dv}. These authors suggested to start from a WZNW model with the
appropriate number of free fermions added, as in the Kazama-Suzuki construction.
Then they gauged the subgroup used by Kazama and Suzuki, allowing for an asymmetry
between the left and right moving sector. Such asymmetric gaugings are severely
constrained by anomaly cancellation conditions and hence the construction of
\cite{Berglund:1995dv} only gives rise to a scarce list of models. If we require
$c_L < 3$, one obtains the $(2,0)$ minimal models of \cite{Berglund:1995dv} in
which the left moving $\mathcal{N}=2$ superconformal algebra or central charge
$c_L = 3k/(k+2)$ is combined with a right moving SU(2) current algebra at level
$k$. In this case, consistency requires $k = 2(Q^2-1)$ so that the lowest allowed
left moving central charge is $c_L = 9/4$.

Quite recently, Gadde and Putrov constructed another infinite family of
$(2,0)$ theories with $c_L = c_R = 3k/(k+2)$, this time for any value of
$k=2,3,\dots$ \cite{Gadde:2016khg}. Once again, the left moving chiral symmetry
is the usual $\mathcal{N} =2$ superconformal algebra, while on the right
their models preserve a subalgebra of the SU(2) current algebra given by
\begin{equation}
 \mathcal{W_R} = \mathrm{SU}(2)_k/\mathrm{U}(1)_{2k} \times
\textrm{U(1)}_{k(k+2)} = \textrm{PF}_k \times \textrm{U}_{k(k+2)}\ ,
\end{equation}
\ie, a product of the parafermionic chiral algebra and a U(1) current
algebra. Let us recall that the sector of parafermions are labeled by
pairs $(l,\alpha)$ with $l=0,1,\dots,k$ and $\alpha=-k+1, \dots, k$ such
that $l+\alpha$ is even. The two pairs $(l,\alpha)$ and $(k-l,k-\alpha)$
correspond to the same sector. The conformal weight of the primary fields
in these sectors satisfies
$$ h^{\textrm{PF}_k}_{(l,\alpha)} =  \frac{l(l+2)}{4(k+2)}
- \frac{\alpha^2}{4k} \textit{ mod } \ 1 \ . $$
The U(1) current algebra $\U(1)_K$, on the other hand, possesses
$K$ sectors with primaries of conformal weight
$$h^{\textrm{U}_{K}}_m = m^2/2K\ .$$
Working with a smaller chiral algebra for right movers, as compared to
the affine current algebra that was used in \cite{Berglund:1995dv}, allows
for additional freedom so that now there is a modular invariant for
any value $k$ of the level. It takes the form
\begin{equation}
\label{Zk}
 Z_k(q,\bar q) = \sum_{l=0}^{k} \sum_{\alpha \in \mathbb{Z}_{2k}}
\sum_{s\in \mathbb{Z}_{k+2}} \chi^{\textrm{SMM}_k}_{(l,2s-\alpha)}(\bar q)
\chi^{\textrm{PF}_k}_{(l,\alpha)}(q)
\chi^{\textrm{U}_{k(k+2)}}_{ks+\alpha}(q)  \ .
\end{equation}
In order to complete the description of these models we also recall that
Neveu-Schwarz sector representations $(l,m)$ of the $\mathcal{N}=2$
superconformal algebra come with $l=0,\dots,k$ and $m=-2k+1, \dots, 2k$
subject to the selection rule $l+m$ even, and field identification
$(l,m) \cong (k-l,2k-m)$. The conformal weight and charge of the
corresponding primaries obey
$$ \bar{h}^{\textrm{SMM}_k}_{(l,m)} = \frac{l(l+2)-m^2}{4(k+2)}
\textit{ mod } 1 \quad , \quad
\bar{q}^{\textrm{SMM}_k}_{(l,m)} = \frac{m}{k+2}
\textit{ mod } 2 \ . $$
The first non-trivial example of the modular invariant \eqref{Zk} appears
for $k=2$ at $c_L = 3/2 = c_R$. It consists of 12 sectors and its modular
invariant reads
\begin{equation}
\label{hetc32}
 Z(q,\bar q) = \sum_{l=0}^{2} \sum_{m=-3}^4
\chi^{\textrm{SMM}_2}_{(l,m)}(\bar q)
\, \chi^{\textrm{Ising}}_l(q) \chi^{\textrm{U}_8}_{m} (q)
  \ .
\end{equation}
Here, Ising stands for the Ising model whose three sectors are labeled
by $l=0,1,2$ and SMM$_2$ denotes the $\mathcal{N}$=2 supersymmetric minimal
model with central charge $c=c_l = 3/2$. Only the six NS sector representation of
the corresponding superconformal algebra appear in the modular invariant.
These are labeled by $(l,m)$ with $l+m$ even and $l=0,1,2$, $m=-3, -2,
\dots, 3,4$. The pairs $(l,m)$ and $(2-l,4-m)$ denote the same representation.

Of course, for each value of the central charge $c_L = c_R = 3k/(k+2)$ one can
also construct a minimal model in which the $(2,0)$ supersymmetry happens to be
extended to $(2,2)$. In particular, for $c_L = c_R= 3/2$ we have at
least two $(2,0)$ models, one heterotic theory with modular invariant
\eqref{hetc32} and the usual diagonal supersymmetric minimal model. In our
numerical analysis below the point $c_L=c_R=3/2$ will appear at the boundary
of the allowed region. We will access this point by studying the four-point
function of an uncharged scalar field $\Phi$ of weight $h = \bar{h}=1/2$. In
both theories, the heterotic and the (2,2) minimal model, this field $\Phi$
involves the same primary $\bar{\varphi}_{(l,m)}= \bar{\varphi}_{(2,0)}$ of
the left moving superconformal algebra. While it is combined with a right
moving primary $\varphi_{(2,0)}$ from the same sector in the $(2,2)$ minimal
model, the right moving contributions in the heterotic theory \eqref{hetc32}
are built as a product of the $l=2$ field $\varepsilon$ in the Ising model
and the identity of the U$_8$ theory, \ie, $\Phi^{\textrm{het}} =
\bar{\varphi}_{(2,0)} \cdot \varepsilon$. Using \eg, a free fermion
representation of the SU(2)$_2$ current algebra it is not difficult to see
that the field $\Phi = \bar{\varphi}_{(2,0)} \cdot \varphi_{(2,0)}$ and
$\Phi^\textrm{het}$ possess identical four point functions. Hence we will
not be able to distinguish between them in our bootstrap analysis below.

Let us also point out that in both models, the field $\Phi$ does not appear 
in the OPE of $\Phi$ with itself. In case of the heterotic theory \eqref{hetc32}, 
this may be seen from the well known fusion rule $\varepsilon \times \varepsilon 
\sim \textit{id}$ of the $c=1/2$ Virasoro algebra. The fusion rules of the  
$\NN =2$ super-Virasoro algebra, which can be found in \cite{Mussardo:1988ck}, 
imply that $\varphi_{(2,0)} \times \varphi_{(2,0)} \sim \textit{id}$ for $k=2$. 
Hence, the field $\Phi$ of the (2,2) minimal model can not appear in its own 
self-OPE, as we had claimed. Our conclusion results from a low level truncation 
in the fusion rules of the $\NN=2$ super-Virasoro algebra.\footnote{Unfortunately, 
this truncation was omitted in \cite{Bobev:2015jxa}.} For values $k\geq 3$, the corresponding 
fields $\Phi$ with left moving quantum numbers $(l,m)=(2,0)$ do appear in their
self-OPE, both for the heterotic and the (2,2) minimal models.

\subsection{The (2,0) crossing equations}
\label{sec:crossing}

The goal of this subsection is to derive the $(2,0)$ crossing symmetry equations \eqref{eq:final_crossing} for
six functions $\hat{g}_i= \hat{g}_i(I_0,\zb), i=1,\dots,6$ of two variables, $I_0$ and $\zb$. In order to do so,
we combine the theory of left moving superblocks from the previous section with the well-known theory of
bosonic blocks for the right movers. Blocks of the latter depend on a single cross ratio $\zb$. So, let us
consider the four-point function of a two-dimensional uncharged superfield, $\Phi (\x,\t,\tb,\xb)$, with
equal holomorphic and anti-holomorphic dimensions  $h=\hb$. Here, $\x$ ($\xb$) denotes the (anti-)holomorphic
bosonic coordinate while $\t$ and $\tb$ are both left-moving (holomorphic) fermionic variables. The four-point
function can then be written as
\be
\label{eq:heteroticfourpt}
\begin{split}
&\langle \Phi(\x_1, \t_1,\tb_1,\xb_1) \Phi(\x_2, \t_2,\tb_2,\xb_2) \Phi(\x_3, \t_3,\tb_3,\xb_3)\Phi(\x_4, \t_4,\tb_4,\xb_4)\rangle
=  \frac{1}{Z_{12}^{2 h}}\frac{1}{Z_{34}^{2 h}}\frac{1}{\xb_{12}^{2\hb}\xb_{34}^{2\hb}} \times  \\[2mm]
& \Big( g_0(I_0,\zb)+ I_1 g_1(I_0,\zb)+I_2 g_2(I_0,\zb)+I_3 g_3(I_0,\zb)+I_4(1-I_0)g_4(I_0,\zb) + I_3 I_4 (1-I_0)g_5(I_0,\zb) \Big)\,.
\end{split}
\ee
This representation of the four-point function follows the one we have used in eqs.\  \eqref{eq:fourpt} and
\eqref{eq:susyexp}, only that now the left moving coordinates are accompanied by right moving bosonic variables
$\xb_i$. Consequently, the functions $f_i(I_0)$ in eq.\ \eqref{eq:fourpt} are replaced by functions $g_i(I_0,\zb)$, containing an additional dependence on the usual cross-ratio
\be
\zb= \frac{\xb_{12}\xb_{34}}{\xb_{13}\xb_{24}}\,, \qquad \xb_{ij}= \xb_i-\xb_j\,.
\ee
All other notations are as in the previous section. Following the usual logic we obtain the crossing equation
by comparing the correlation function \eqref{eq:heteroticfourpt} with the one in the crossed channel in which
$(x_1,\t_1,\tb_1,\xb_1)$ and  $(x_3, \t_3,\tb_3,\xb_3)$ are exchanged. This leads to the equation
\begin{align}
\begin{split}
&g_0(I_0,\zb)+ I_1 g_1(I_0,\zb)+I_2 g_2(I_0,\zb)+I_3 g_3(I_0)+I_4 g_4(I_0,\zb) +I_3 I_4 (1-I_0)g_5(I_0,\zb) =
\\[2mm]
& (I_0+I_1)^{2 h} \left(\frac{\zb}{\zb-1}\right)^{2\hb} \Bigg( g_0(I_0^{t},1-\zb) +
\frac{-I_1}{I_0 + I_1}g_1(I_0^{t},1-\zb)+ \frac{2I_4(1-I_0)-I_2}{I_0+I_1} g_2(I_0^{t},1-\zb) \\[2mm]
&+ \frac{I_4(1 - I_0) + I_3 - I_2}{I_0} g_3(I_0^{t},1-\zb)- I_4 g_4(I_0^{t},1-\zb) +
\frac{I_3I_4(1 - I_0) }{I_0(I_0+I_1)}g_5(I_0^{t},1-\zb)\Bigg)\,.
\end{split}
\label{eq:crossing}
\end{align}
Upon swapping $(x_1,\t_1,\tb_1)$ with $(x_3, \t_3,\tb_3)$ the invariants $I_i$ become $I_i^t$. The
latter may be expressed in terms of $I_i$ as
\begin{align}
I_0^{t} = \frac{1+I_1}{I_0+I_1} \,, \quad  I_1^{t} & = \frac{-I_1}{I_0+I_1} \,, \quad
I_2^{t}= \frac{2I_4(1-I_0)-I_2}{I_0+I_1}\,, \quad I_3^{t}= \frac{I_4(1-I_0)+I_3-I_2}{I_0}\,,  \nn\\[2mm]
 I_4^{t} & =-I_4\,, \qquad I_3^{t}I_4^{t}(1-I_0^{t})=\frac{I_3 I_4 (1-I_0)}{I_0 (I_0+I_1)}\,.
\end{align}
Next we Taylor expand equation \eqref{eq:crossing} in the nilpotent invariants ($I_{i \neq 0}$)
with the end result collected in eq.\ \eqref{eq:crossing_expanded} as it is rather long.
By comparing the coefficients of the six different nilpotent structures we obtain a system
of six crossing equations for the six functions $g_i = g_i(I_0,\zb), i=0, \dots 5,$ of the
two variables $I_0$ and $\zb$.

Finally we want each function of the two cross-ratios $I_0$ and $\zb$ to admit a block decomposition that can be interpreted as the exchange of a given representation in the correlation function
of the various operators that make up the external superfield $\Phi(\x_1,\t_1,\tb_1,\xb_1)$.
This is achieved as in eq.\  \eqref{eq:ftofhat} by going to  a ``primary basis'' which can be decomposed in terms of the $\hat{f}_i$ blocks we have determined, \ie, we rewrite the crossing
equations in terms of $\hat{g}_i(I_0,\zb)$, where the $\hat{g}_i$ are related to the $g_i$ in the same way the $\hat{f}_i$ are related to $f_i$. In addition, we express the variable $I_0$ in terms
of a new variable
\be
\bzb=\frac{I_0}{I_0-1}\,,
\ee
that reduces to the standard cross ratio $z$ upon setting all the fermionic variables to zero.
With these notations, the six crossing equations can be written in the form
\begingroup
\allowdisplaybreaks[1]
\begin{align}
\begin{split}
 0 =\ & (1-z)^{2 h} \hat{g}_0(\bzb ,\zb)-z^{2 h} \hat{g}_0(1-\bzb ,1-\zb)\,, \\[2mm]
 0 =\ & (1-z)^{2 h+1} \hat{g}_3(\bzb ,\zb)-z^{2 h+1} \hat{g}_3(1-\bzb ,1-\zb)\,, \\[2mm]
 0 =\ & -2 (z-1) z \left((z-1) z^{2 h} \hat{g}_0(\bzb ,\zb){}^{(1,0)}(1-\bzb ,1-\zb)+z (1-z)^{2 h} \hat{g}_0(\bzb
   ,\zb){}^{(1,0)}(\bzb ,\zb)\right)\,,  \\[2mm]
   & +2 z^{2 h+1} \hat{g}_1(1-\bzb ,1-\zb)-2 (1-z)^{2 h+1} \hat{g}_1(\bzb
   ,\zb)\,,\\[2mm]
0 =\ & (1-z)^{2 h+1} \hat{g}_3(\bzb ,\zb)+z^{2 h+1} \hat{g}_3(1-\bzb ,1-\zb)\,,\\[2mm]
0 =\ & (1-z)^{2 h+1} \hat{g}_4(\bzb ,\zb)+z^{2 h+1} \hat{g}_4(1-\bzb ,1-\zb)\,,\\[2mm]
0 =\ & 2 h (z-1) z^{2 h+1} \hat{g}_0(\bzb ,\zb){}^{(1,0)}(1-\bzb ,1-\zb)+(z-1) z^{2 h+2} \hat{g}_1(\bzb
   ,\zb){}^{(1,0)}(1-\bzb ,1-\zb) \\[2mm]
& +z (1-z)^{2 (h+1)} \hat{g}_1(\bzb ,\zb){}^{(1,0)}(\bzb ,\zb)+2 h z
   (1-z)^{2 h+1} \hat{g}_0(\bzb ,\zb){}^{(1,0)}(\bzb ,\zb) \\[2mm]
   & +2 h^2 (2 z-1) z^{2 h} \hat{g}_0(1-\bzb ,1-\zb)+2 h^2
   (2 z-1) (1-z)^{2 h} \hat{g}_0(\bzb ,\zb)  \\[2mm]
   & +z^{2 h+1} (2 h (z+2)+z) \hat{g}_1(1-\bzb ,1-\zb)+z^{2 h+2} \hat{g}_5(1-\bzb
   ,1-\zb)  \\[2mm]
   & -(1-z)^{2 (h+1)} \hat{g}_5(\bzb ,\zb)+(1-z)^{2 h+1} (2 h (z-3)+z-1)  \hat{g}_1(\bzb ,\zb)\,.
\label{eq:final_crossing}
\end{split}
\end{align}
Note that we have written the equations in a way such that they have an obvious symmetry under ${\bf z} \to 1-{\bf z}$ and $\zb \to 1-\zb$. This will prove to be convenient for the numerical implementation.
\endgroup 

\subsection{Block expansions}
\label{sec:heteroticblockexpansion}

Each of the functions $\hat{g}_i$ in the crossing equation \eqref{eq:final_crossing} admits a
decomposition into superblocks in the left moving variable ${\bf z}$ and regular bosonic blocks
depending on $\zb$
\be
\hat{g}_i({\bf z}, \zb) = \sum_{h_{\ex},q_{\ex},\hb_{\ex}}
\lambda_{h_{\ex},q_{\ex},\hb_{\ex}}^2 \hat{f}_i({\bf z}) \gsl_{\hb_{\ex}} (\zb) \,.
\label{eq:blockdec}
\ee
We recall that on the supersymmetric (left) side, $h_{\ex},q_{\ex}$ are the quantum numbers of
the superconformal primary in a given supermultiplet, even if the operator appearing in the OPE
is not the superprimary itself.
Unitarity requires that the summation in eq.\ \eqref{eq:blockdec} is restricted by $h_{\ex}
\geqslant \frac{q_{\ex}}{2}$ and $\hb_{\ex} \geqslant 0$.
Here, the superblocks are given by eq.\  \eqref{eq:hatf_ans}, with the coefficients fixed by eqs.\ \eqref{eq:atypesol},
\eqref{eq:btypesol} and \eqref{eq:ctypesol}. Recall that since we are considering identical external fields, Bose symmetry fixes all coefficients as given in eq.\  \eqref{eq:identicalops}, up to a normalization. We normalize them by setting $X_0=1$, with
$X=a,b,c$ depending on which of the solutions we consider.
The bosonic blocks, on the other hand, possess the standard expression
\be
\gsl_{\hb_{\ex}} (\zb) = \zb^{\hb} {}_2 F_1  \left(\hb_{\ex}, \hb_{\ex}, 2\hb_{\ex}, \zb \right)\ .
\ee
With our normalizations, the squares $\lambda_{h_{\ex},q_{\ex},\hb_{\ex}}^2$ of the OPE coefficients are the same that would appear
in the four-point functions of the superconformal primary of $\Phi (\x,\t,\tb,\xb)$, \ie, when
we set all fermionic variables in eq.\ \eqref{eq:heteroticfourpt} to zero. Hence, they are
positive numbers.

On the supersymmetric side, we found in section \ref{sec:longblockssol}, that there could be
the following types of operators exchanged
\begin{itemize}
\item The superconformal primary (of dimension $h_{\ex}$) of an uncharged ($q_{\ex}=0$) superconformal multiplet is exchanged -- the solution given by the $a_i$ in eq.\ \eqref{eq:atypesol},
\item The superconformal descendant of dimension $h_{\ex}+1$ of an uncharged ($q_{\ex}=0$) superconformal multiplet whose superconformal primary has dimension $h_{\ex}$ -- the solution given by $b_i$ in eq.\ \eqref{eq:btypesol},
\item The uncharged superconformal descendant of dimension $h_{\ex}+\tfrac{1}{2}$ of a charged superconformal multiplet whose superconformal primary has dimension $h_{\ex}$ and charge $q_{\ex}=\pm1$ --  the solution given by $c_i$ in eq. \eqref{eq:ctypesol} with $q_{\ex}=\pm 1$.
\end{itemize}
Now we want to see which pairings of the above quantum numbers with the anti-holomorphic dimension $\hb$ can appear in the OPE of identical uncharged scalars.
Defining
\be
\Delta= h + \hb\,, \qquad \ell=h - \hb\,,
\ee
we want to obtain the range of $\Delta_{\ex}$ and $\ell_{\ex}$ for operators that can appear in the self-OPE of the external superfield.
Note that in two-dimensions, since the conformal group factorizes, parity does not exchange states in the same representation. In particular $\ell$ can be both positive and negative, and since we focus on $\NN=(2,0)$ theories (which clearly have no symmetry between ${\bf z}$ and $\zb$, as visible in eq.\ \eqref{eq:blockdec}), we must consider both signs of $\ell$ independently.\footnote{For the bosonic case, when putting together holomorphic blocks to make the whole conformal block, one usually symmetrizes in $z \leftrightarrow \zb$,  and therefore can restrict the OPE decompositions to positive spin (see \eg, \cite{Osborn:2012vt}). Parity odd blocks, anti-symmetric under this exchange, were considered in \cite{Heemskerk:2010ty}.} However $\ell$ should still be half-integer for single-valuedness of correlation functions.
This means that the sum \eqref{eq:blockdec} will have a discrete parameter $\ell_{ex}$, and a continuous one $\Delta_{\ex}$ satisfying
\be
\Delta_{ex} \geqslant |\ell_{\ex} |\,, \;\; \mathrm{for}\; q_{\ex}=0\,, \quad
\Delta_{ex} \geqslant \ell_{\ex} \,, \;\; \mathrm{for}\; |q_{\ex}| \leqslant 2\ell_{\ex}\,, \quad
\Delta_{ex} \geqslant |q_{\ex} | - \ell_{\ex} \,, \;\; \mathrm{for}\; |q_{\ex}| \geqslant 2\ell_{\ex}\,.
\ee
Furthermore, Bose symmetry constrains the spin of the operators, appearing in the OPE of the superconformal primary of $\Phi(\x,\t,\tb,\xb)$, to be even, putting constraints on the spin of the superconformal primary of multiplet $\ell_{\ex}$.

Of the multiplets appearing in the OPE three are noteworthy. One corresponds to the identity operator, which has $\Delta_{ex}=\ell_{ex}=q_{\ex}=0$ and comes from the $a_i$ solution in eq.\ \eqref{eq:atypesol}. The other two correspond to the holomorphic and anti-holomorphic stress tensors, which are given respectively by a $b_i$ solution in eq.\ \eqref{eq:btypesol} with $\Delta_{\ex}=\ell_{\ex}=1$, $q_{\ex}=0$; and by an $a_i$ solution with $\Delta_{ex}=-\ell_{ex}=2$ and $q_{\ex}=0$.


\subsection{Numerical implementation}
\label{sec:numericalimplementation}

To analyze the crossing equations \eqref{eq:final_crossing} we proceed numerically, as pioneered in \cite{Rattazzi:2008pe}, using the SDPB solver of  \cite{Simmons-Duffin:2015qma}. We follow the, by now standard, procedure to obtain numerical bounds (see, \eg, \cite{Rychkov:2016iqz, Simmons-Duffin:2016gjk} for reviews).
In the block decomposition \eqref{eq:blockdec} we approximate the superblocks by polynomials in the exchanged dimension $\Delta_{\ex}$, as first implemented in \cite{Poland:2011ey}, and truncate the infinite sum over the spins from $- L_{max} \leqslant \ell \leqslant L_{max}$.\footnote{Note that due to the explicit $\Delta_{\ex}$ factors in the crossing equations, and derivatives of blocks, one must be careful to consistently approximate all terms in the crossing equation to a polynomial of the same degree in $\Delta_{\ex}$.}

By searching for six-dimensional linear functionals
\begin{equation}
\label{eq:functional}
\vec{\Phi} = \sum_{n,m=0}^{n+m \leqslant \Lambda} \vec{\Phi}_{m,n} \partial_{\bf z}^m \partial_{\zb}^n \vert_{{\bf z}=\zb=\tfrac{1}{2}}\,,
\end{equation}
whose action on the crossing equations is subject to a given set of conditions, we can rule out assumptions on the spectrum of operators $\{ \Delta_{\ex}, \ell_{\ex}, q_{\ex}\}$ appearing in the OPE, and on their OPE coefficients.
The cutoff $\Lambda$ implies we are effectively studying a Taylor series expansion of the crossing equations, truncated by $\Lambda$. Therefore for each $\Lambda$, we obtain valid bounds, that will get stronger as we increase the number of terms kept in the Taylor expansion.
Each of the equations \eqref{eq:final_crossing} has a definite symmetry under ${\bf z} \to 1-{\bf z}$ and $\zb \to 1-\zb$, according to which only even or odd $m+n$ derivatives in eq.\ \eqref{eq:functional} will be non-trivial. However, unlike the typical bootstrap setups, the equations have no symmetry in $z \leftrightarrow \zb$ and we cannot restrict to derivatives with $m < n$.

\subsection{\texorpdfstring{Numerical results for $\NN=(2,0)$ theories}{Numerical results for N=(2,0) theories}}
\label{sec:results}

\subsubsection{Central charge bounds}
\label{sec:numericalcbounds}

In exploring the space of $\NN=(2,0)$ SCFTs the first question one wants to answer 
concerns the range of allowed central charges.
Here we explore what values are allowed for both left and right central charges, while allowing the other one to be arbitrary, and compare the numerical bounds with the known landscape of theories described in section \ref{sec:landscape}.
A peculiarity of two dimensions, as already discussed in \cite{Vichi:2011zza, El-Showk:2014dwa}, is that one cannot find a lower bound on the central charge without imposing a small gap in the spectrum of scalar operators. Therefore to obtain central charge bounds we require that all scalar superprimaries appearing in the OPE of our external field have dimension larger than a certain value, which we denote by $h_{\gap}= \hb_{\gap}$. The bounds are then obtained for various different values of $h_{\gap}= \hb_{\gap}$.\footnote{That it is necessary to impose a gap is expected from the fact that the unitarity bounds in two dimensions do not have a gap between the dimension of the identity operator ($0$) and that of the first generic operator. This implies that the optimization problem we try to solve, by minimizing the value of the functional on the identity, while remaining positive on all other blocks, is only possible if the continuum is isolated from the identity by a gap imposed by hand. Why the gap must be at least of the order of $h_{\gap} \sim 1.7 h$, as empirically observed in the numerical results, is not clear to us.}

\subsubsection*{Right central charge}

\begin{figure}[htb]
             \begin{center}
              \includegraphics[scale=0.6]{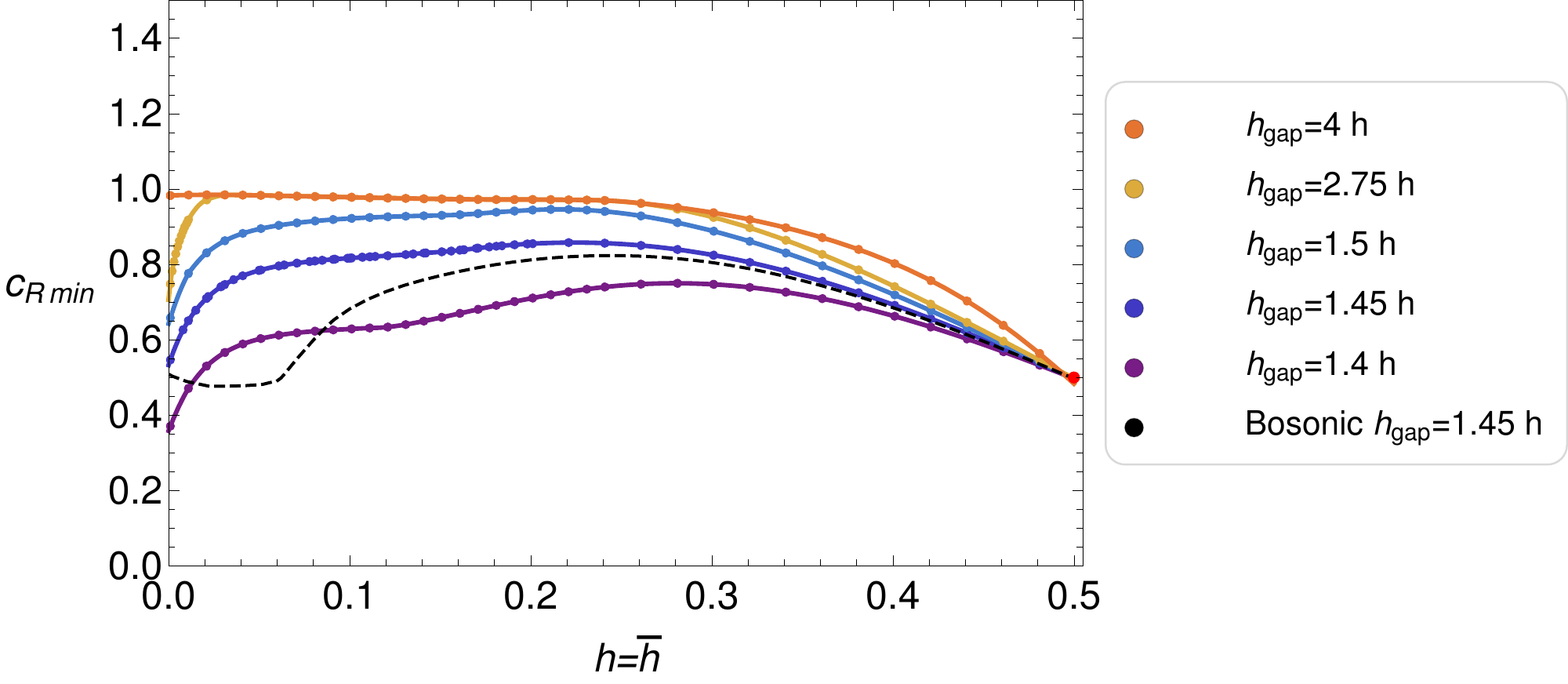}
                        \caption{Lower bound on the allowed right central charge $c_R$ (non-supersymmetric side) of $\NN=(2,0)$ SCFTs as a function of the external dimension, $h=\hb$, after imposing different gaps on the spectrum of superprimary scalar operators $h_{\gap}=\hb_{\gap}$. The lines with dots correspond to the full set of crossing equations. The dashed black line corresponds to the bound obtained from the crossing equations of the superconformal primary alone, which matches with the bosonic bootstrap bounds, and is obtained for a single $h_{\gap}=1.45 h$.  The red dot marks the central charge and external dimension of the known $(2,2)$ and $(2,0)$ minimal models described  below eq.\  \eqref{hetc32} (see text for discussion). The bounds were obtained for $\Lambda=20$ which, with the shown scale is enough to have obtained a converged plot as exemplified in figure \ref{Fig:convergence}.}
             \label{Fig:crbound}
            \end{center}
\end{figure}

We start by obtaining a lower bound on $c_R$ (the central charge of the non-supersymmetric side), displayed on figure \ref{Fig:crbound}, overlapped with the bound obtained from the purely bosonic crossing equations.
The bounds in figure \ref{Fig:crbound} obtained for the full set of crossing equations \eqref{eq:final_crossing} (colored dots and lines) assume various different values of $h_{\gap}$, while for the crossing equations of just the superconformal primary (the first equation in \eqref{eq:final_crossing}) we picked a single illustrative value of $h_{\gap}$ (dashed black line).\footnote{The  first equation in the list \eqref{eq:final_crossing} is exactly the equation for a bosonic theory, and bounds for various gaps were obtained in \cite{Vichi:2011zza}.}
In order to obtain a non-trivial central charge bound we found we needed to impose a gap in the scalar superprimary spectrum of $h_{\gap} =\hb_{\gap} \sim 1.7 h$, where $h=\hb$ is the dimension of the external superprimary. The size of the minimum gap appears to be similar to the one needed in \cite{Vichi:2011zza} for the bosonic case. The bounds are shown only for $\Lambda=20$, to avoid cluttering, which is enough for them to have approximately converged in the scaled used. (The rate of convergence is exemplified for the left central charge bound $c_L$ in figure \ref{Fig:convergence}.) Finally, note that the bounds start with $h=\hb$ slightly above zero, as at the point $h=\hb=0$ the external field becomes shorter (it becomes the identity) and the blocks we computed are not valid.

The bounds for the full set of crossing equations are much stronger than the purely bosonic ones, in particular the minimum corresponding to the central charge of the two-dimensional Ising model is absent.
This exemplifies the amount of constraints lost if one were to consider only the correlation function of the superconformal primary, for which the ``superblocks'' are just bosonic conformal blocks. Even though Bose symmetry fixes the four-point function of external superdescendants in terms of that of the external superprimary one, the crossing equations for external descendants provide non-trivial constraints, further reducing the space of allowed CFTs. This is stark contrast with the case of half-BPS operators, such as the two-dimensional chiral operators considered in \cite{Bobev:2015jxa}, where the only invariants were the supersymmetrization of the bosonic cross-ratios $u$ and $v$.

We see the bounds exhibit a strong dependence on the gap imposed, with the exception of a neighborhood of $h=\hb=\tfrac{1}{2}$, where all bounds appear to give the same value approaching $c_R=\tfrac{1}{2}$.
This leads to the natural question of whether there is a physical theory with $h=\hb=\tfrac{1}{2}$ saturating our bounds. Looking at the landscape of known physical $\NN=(2,0)$ SCFTs, briefly described in section \ref{sec:landscape}, we see that the uncharged scalar operators of most of the $\NN=(2,0)$ models there described, and also of the $\NN=(2,2)$ minimal models, have the property that said scalar appears in its self-OPE. Therefore by imposing a gap $h_{\gap} \geqslant 1.7 h$ we exclude all these theories by hand. The exceptions are the $\NN=(2,2)$ minimal model with central charge $c_L=c_R=\tfrac{3}{2}$, and the heterotic model described around eq.\ \eqref{hetc32}. Both these theories have an uncharged scalar of dimension $h=\hb=\tfrac{1}{2}$ and thus should be allowed in our setup.
To understand how these theories should appear in our plots we must first point out that by $c_R$ we mean the central charge coefficient read off from the exchange of a superprimary operator on the left, and $\sl(2)$ primary on the right, with $h_{\ex}=q_{\ex}=0$ and $\hb_{\ex}=2$. It could happen that there is more than one such operator. For example, if the theory actually has $\NN=2$ supersymmetry on the right side, we expect there to be two such operators: the anti-holomorphic stress tensor, and the Sugawara stress tensor, made out of the $\U(1)$ current. As such we are not guaranteed to be bounding the OPE coefficient of the anti-holomorphic stress tensor.

Let us start by describing the $\NN=(2,2)$ minimal model for which the $h=\bar{h}=\tfrac{1}{2}$ operator of charge zero does not appear in its self-OPE, and thus should appear inside our allowed region.\footnote{This minimal model also has chiral operators ($h=\tfrac{|q|}{2}$) of charge $\pm \tfrac{1}{4}$ and $\pm \tfrac{1}{2}$, which appear inside the dimension bounds of Figure~5 of \cite{Bobev:2015jxa} (the $k=2$ minimal model), and at least for the $\Lambda$ considered there, not saturating them.} As pointed out above we are not guaranteed to be bounding $c_R$, which for this model should be $\tfrac{3}{2}$. In this case we are obtaining a sum of OPE coefficient squared, namely that of the stress tensor and of the Sugawara stress tensor. Computing this OPE coefficient we find it should give rise to an apparent central charge of $c_R=\tfrac{1}{2}$, and thus this solution appears to saturate our bounds, and is indicated by a red dot in figure \ref{Fig:crbound}.
Next we turn to the heterotic model described around eq.\ \eqref{hetc32}. In this case there is no supersymmetry on the right side, and the coefficient we are bounding corresponds exactly to $c_R$. Again in this case $c_R=\tfrac{1}{2}$, which is indicated by the same red dot in figure \ref{Fig:crbound}, and this solution too appears to saturate our bounds. As explained in section the correlation function of these two solutions are the same, and this is not in conflict the fact that there is a unique solution to the crossing equations for theories that sit on the numerical exclusion curves \cite{Poland:2010wg,ElShowk:2012hu}.

\subsubsection*{Left central charge}

\begin{figure}[htb]
             \begin{center}
              \includegraphics[scale=0.6]{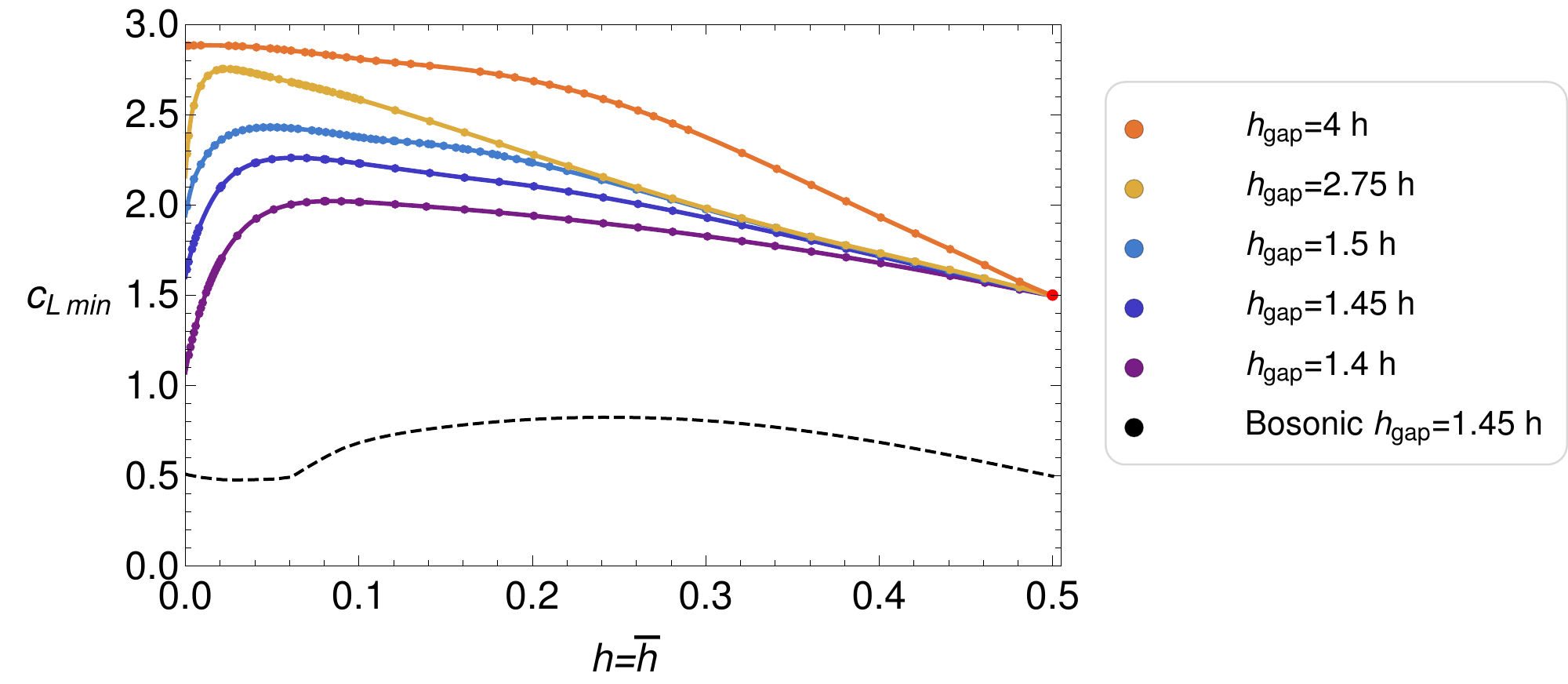}
                        \caption{Lower bound on the allowed left central charge $c_L$  (supersymmetric side) of $\NN=(2,0)$ SCFTs as a function of the external dimension, $h=\hb$, after imposing different gaps on the spectrum of superprimary scalar operators $h_{\gap}=\hb_{\gap}$. The lines with dots correspond to the full set of crossing equations. The dashed black line corresponds to the bound obtained from the crossing equations of the superconformal primary alone, which matches with the bosonic bootstrap bounds, and is shown for a single value of $h_{\gap}$. The red dot marks the central charge and external dimension of the known $(2,2)$ and $(2,0)$ minimal models discussed below eq.\  \eqref{hetc32} which have identical four-point functions for this external operator. The bounds were obtained for $\Lambda=20$, and the rate of convergence of the numerical bounds is shown in figure \ref{Fig:convergence} for $h=\hb=0.5$.}
             \label{Fig:clbound}
            \end{center}
\end{figure}

\begin{figure}[htb]
             \begin{center}
              \includegraphics[scale=0.35]{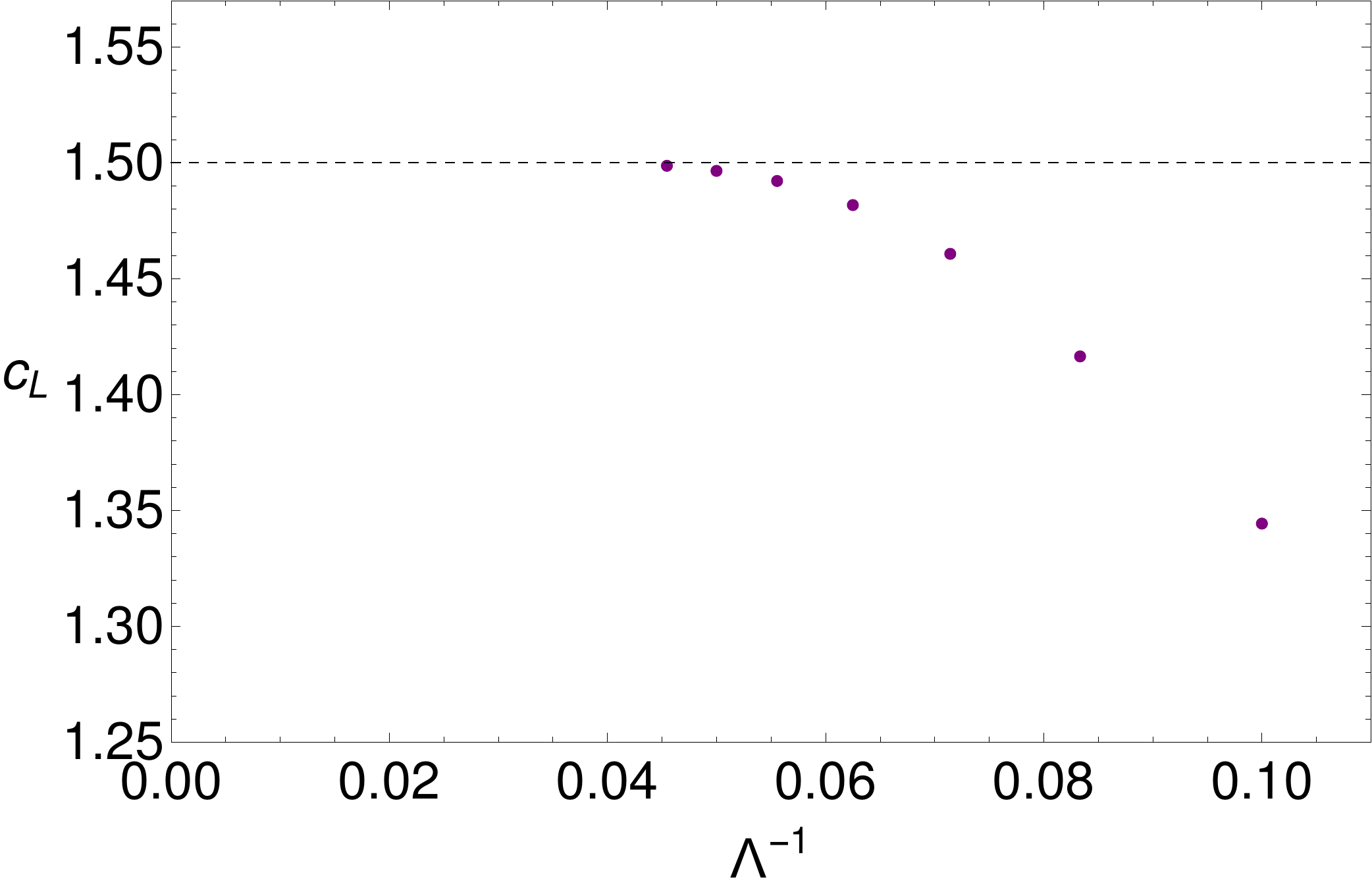}
                        \caption{ Lower bound on the allowed left central charge $c_L$  (supersymmetric side) of $\NN=(2,0)$ SCFTs for $h=\hb=0.5$ and $h_{\gap}=\hb_{\gap}=0.8$, as a function of the inverse of the number of derivatives ($\Lambda^{-1}$) to exemplify the convergence of our numerical results with $\Lambda$.}
             \label{Fig:convergence}
            \end{center}
\end{figure}

Next we turn to $c_L$ (the central charge of the $\NN=2$ side) shown in figure \ref{Fig:clbound}, where we obtain, as expected, a much stronger bound than for the non-supersymmetric side.
We obtain $c_L$ from the OPE coefficient of the exchange of the holomorphic stress tensor, which is a global superdescendant of the $\U(1)$ current. Thus, unlike in the $c_R$ case, the stress tensor is distinguished from the Sugawara stress tensor: the latter is part of a global superprimary, while the former is always a superdescendant.
As before to obtain a non-trivial $c_L$ bound we must impose a gap at least of the order of $1.7 h$. The plot is obtained at fixed $\Lambda=20$ and, once again, we mark the position of the $\NN=(2,2)$ minimal model and the heterotic model described in section \ref{sec:landscape} as a red dot.
Again, while the bounds display a large dependence on the gap imposed, for external dimension $h=\hb=\tfrac{1}{2}$ all gaps give the same bound, around $\tfrac{3}{2}$. The dependence of the bounds on the cutoff $\Lambda$ is  shown for this value of the external dimension in figure \ref{Fig:convergence}, where we see for $\Lambda=20$ the bounds have almost stabilized to a value close to $\tfrac{3}{2}$.  This is precisely the central charge of the $\NN=(2,2)$ minimal model and the heterotic model, which appear to also saturate both the  $c_R$ and $c_L$ bounds.
Recall from section \ref{sec:landscape} that the four-point function of both this models is equal, corresponding to the unique solution obtained when a bound is saturated.

Finally, if we impose that imposing $c_L=c_R=c$ we seem to find a bound on $c$ identical to that of figure \ref{Fig:clbound}, for the cases of $h_{\gap}=\hb_{\gap}$ we tested. This follows from a technical subtlety, namely, the functional is normalized to one on the sum of the holomorphic and anti-holomorphic stress tensor blocks, but this allows it to be zero on one of them, and one on the other. As such we are obtaining a bound on the minimum of both OPE coefficients, which are inversely proportional to the central charge, and thus what we obtain is the maximum of the $c_L$ and $c_R$ bounds, explaining the observed feature.

\subsubsection{Dimension bounds}
\label{sec:dimbounds}

\begin{figure}[htbp!]
             \begin{center}
              \includegraphics[scale=0.4]{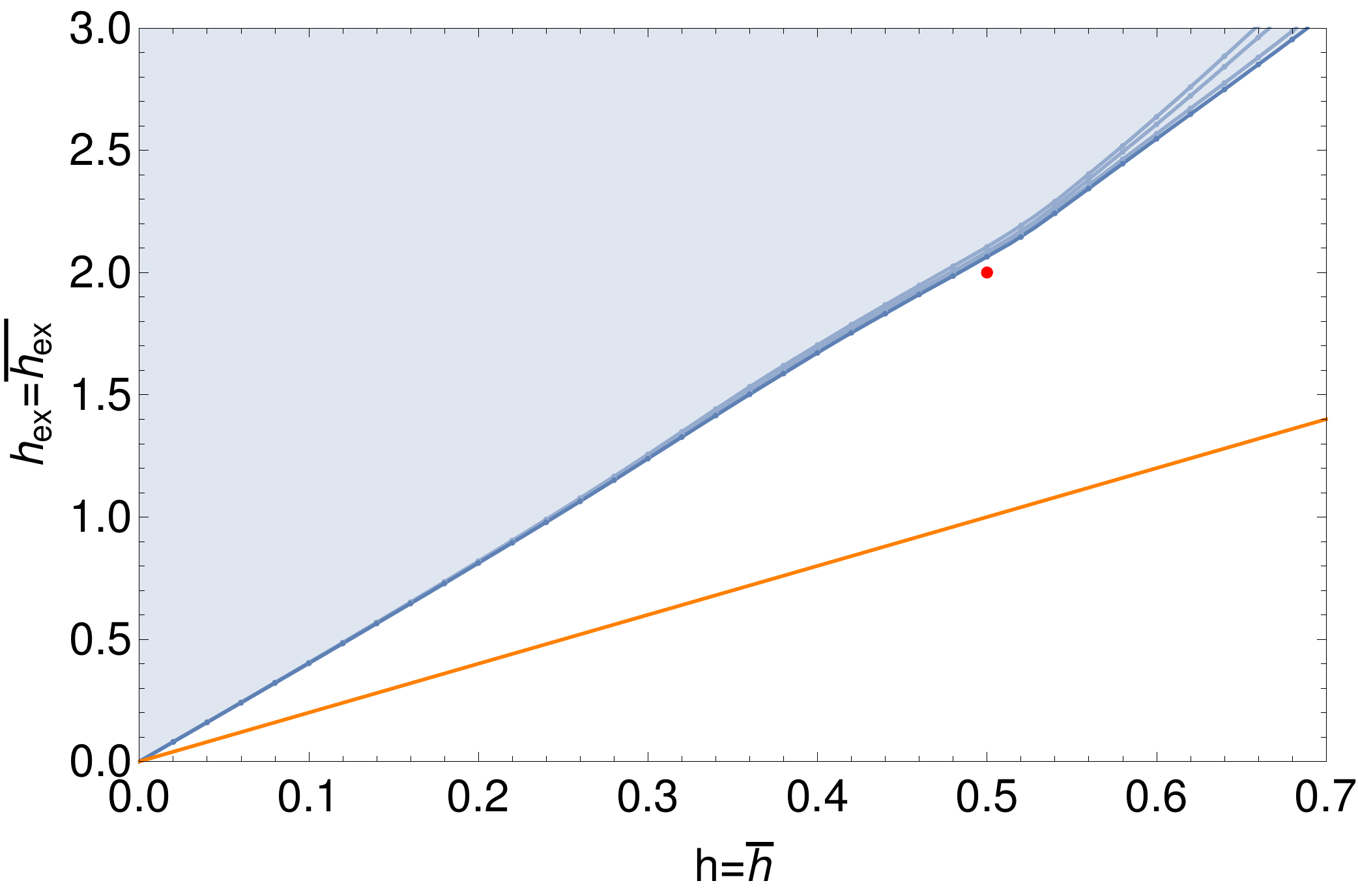}
                        \caption{Upper bound on the dimension of the first uncharged scalar long superconformal primary that appears in the OPE, as a function of the dimension of the external operator $h=\hb$ for $\Lambda=16,18,\ldots, 22$ derivatives.  The red dot marks the dimension of the known $(2,2)$ and $(2,0)$ minimal models discussed below eq. \eqref{hetc32}. The orange line corresponds to the generalized free field theory solution $h_{\ex}=2 h$.
                      }
             \label{Fig:dimbound}
            \end{center}
\end{figure}

Lastly, we turn to bounding the dimensions of the first long scalar operators, whose global superconformal primaries appear in this OPE (this corresponds to the $a_i$ solution in eq.\ \eqref{eq:atypesol}. The upper bound on the dimension of the superconformal primary is shown in figure \ref{Fig:dimbound} for various values of the cutoff $\Lambda$. The orange line in the plot corresponds to the solution of generalized free field theory, \ie, the four-point function given by a sum of products of two-point functions $h_{\ex}=2 h$.

Note that since we are only using \emph{global} superconformal blocks and not Virasoro superblocks we should expect super Virasoro descendants to appear in the OPE independently of their superconformal primaries. In particular the following descendant of the identity $\left(J_{-1} J_{-1}-\tfrac{2}{3}L_{-2}\right) \bar{L}_{-2} |0\rangle$ corresponds to a scalar operator of dimension $h_{\ex}=\hb_{\ex}=2$, which is a global superprimary, and therefore should appear in the OPE channel we are studying in figure \ref{Fig:dimbound}.
For $h \lesssim 0.5$ the numerical results demand an operator of a smaller dimension to be present, while for $h \gtrsim 0.5$ the numerical bounds allow for solutions without such $h_{\ex}=\hb_{\ex}=2$ operators.

The four-point function of the $\NN=(2,2)$ and the $\NN=(2,0)$ models discussed above only has one super Virasoro multiplet being exchanged, that of the vacuum. Therefore we expect the dimension of the first global superconformal primary to be exactly $h_{\ex}=\hb_{\ex}=2$ in both cases, marked by the red dot in figure \ref{Fig:dimbound}. The numerical upper bound on the dimension is converging slower than the central charge bounds (figure \ref{Fig:convergence}), and it is not clear whether it is converging to the red dot, although it seems plausible.

The remaining $\NN=(2,2)$ minimal models, and $\NN=(2,0)$ theories described in subsection \ref{sec:landscape} share the property that the external field appears in its own OPE, \ie, their solution corresponds to $h_{\ex}=\hb_{\ex}=h=\hb$, for $h < \tfrac{1}{2}$. This means they are deep inside the allowed region in figure \ref{Fig:dimbound}, below the generalized free field theory solution (orange line in the plot). This leaves open the question  of whether there exist new theories saturating the numerical bounds for $h < 0.5$, or if the solution to crossing symmetry of this particular correlator cannot be part of a full-fledged SCFT.

One could hope that by allowing the external field to appear in its own OPE, the remaining minimal models would also saturate the numerical bounds. However the next scalar in the minimal models also sits well inside the numerical bound of figure \ref{Fig:dimbound}, and since we cannot force the external scalar to be exchanged, only to allow for its presence, we end up with the same result as in figure \ref{Fig:dimbound}. We could keep repeating the procedure, allowing for both the external scalar, and the first operator exchanged after it in the known solutions. However, preliminary explorations suggest the resulting bound would be very weak.  
\section{Consequences for four-dimensional physics}
\label{sec:Neq3}

Finally, we discuss the implications of the blocks we computed in section \ref{sec:superblocks} for four-dimensional $\NN=3$ SCFTs.
It was shown in \cite{Beem:2013sza} that any $\NN \geqslant 2$ SCFT in four dimensions admits a subsector isomorphic to a two-dimensional chiral algebra. Here we only briefly describe the construction and refer to \cite{Beem:2013sza} for all the details.
The chiral algebra is obtained by restricting operators to lie on a plane, on which we put coordinates $(z,\zb)$, and passing to the cohomology of a certain nilpotent supercharge $\qq$, that is a linear combination of a Poincar\'{e} and a conformal supercharge.
The anti-holomorphic dependence is $\qq$ exact, and cohomology classes of local operators correspond to meromorphic operators on the two-dimensional plane on which we restricted the operators to lie.
We call operators in the cohomology of said supercharge ``Schur operators'', since they correspond precisely to the class of operators contributing to the Schur limit of the superconformal index \cite{Kinney:2005ej,Gadde:2011uv,Rastelli:2014jja}.
The stress tensor multiplet (denoted by $\hat{\CC}_{0,(0,0)}$ in the notation of \cite{Dolan:2002zh}) of an $\NN=2$ SCFT contains one Schur operator, giving rise, in the cohomology, to a two-dimensional operator acting as the meromorphic stress tensor $T(z)$.\footnote{Note that in this section we are using $z$ instead of $x$ for the holomorphic coordinate.} Therefore, the global $\sl(2)$ symmetry on the plane is enhanced to the full Virasoro algebra, with the two-dimensional central charge determined in terms of the four-dimensional $c$ anomaly coefficient,
\be
c_{2d}= - 12 c_{4d}\,.
\ee
Similarly, global symmetries of the four-dimensional theory give rise to affine Kac-Moody current algebras, with level determined from the four-dimensional flavor current central charge
\be
k_{2d}=-\frac{k_{4d}}{2}\,.
\ee
The two-dimensional affine current $J(z)$ arises from a Schur operator in the four-dimensional $\hat{\BB}_1$ multiplet, that also contains the conserved flavor symmetry current.
More generally, each $\NN=2$ superconformal multiplet contributes at most one (non-trivial) operator to the cohomology, giving rise in two dimensions to global
$\sl(2)$ primaries.
This construction uses up all of the supersymmetry of a pure $\NN=2$ theory, and the two-dimensional chiral algebra has no supersymmetry left. However, if the four-dimensional theory has more supersymmetry, then the chiral algebra will also be supersymmetric. This follows immediately from the fact that the extra supercharges, enhancing the supersymmetry beyond $\NN=2$, commute with $\qq$ and thus relate different representatives of $\NN=2$ multiplets in cohomology.
This is the case of theories with $\NN=4$ supersymmetry, for which the chiral algebra will necessarily contain the ``small'' $\NN=4$ super algebra, as discussed in detail in \cite{Beem:2013sza}.
If the theory has instead $\NN=3$ supersymmetry one will end up precisely with a $\NN=2$ two-dimensional chiral algebra as first discussed in \cite{Nishinaka:2016hbw}, with the full list of $\NN=3$ supermultiplets containing Schur operators given in \cite{Lemos:2016xke}.

\subsection{\texorpdfstring{Four-dimensional $\NN=3$ SCFTs}{Four-dimensional N=3 SCFTs}}

The first examples of pure $\NN=3$ SCFTs (\ie, theories which do not have $\NN=4$ supersymmetry) were recently constructed using a generalization of orientifolds in string theory, called S-folds, in \cite{Garcia-Etxebarria:2015wns}.\footnote{Already in \cite{Ferrara:1998zt} a truncation of type IIB supergravity, whose CFT dual would correspond to a
 four-dimensional $\NN=3$ SCFT, had been considered.}
Several properties of pure $\NN=3$ SCFTs can be obtained from representation theory alone, which had been studied long ago in \cite{Dobrev:1985qv,Minwalla:1997ka,Kinney:2005ej}, but only recently was the case of $\NN=3$ explored in detail \cite{Aharony:2015oyb}, shortly before the first $\NN=3$ theories were constructed. Similarly to the $\NN=4$ case, the $a$ and $c$ conformal anomalies of $\NN=3$ SCFTs have to be equal, and pure $\NN=3$ theories cannot have any flavor symmetry which is not an R-symmetry.
They are also isolated theories, in the sense that pure $\NN=3$ theories have no exactly marginal deformations.\footnote{The only $\NN=3$ superconformal multiplet which could accommodate supersymmetric exactly marginal deformations also contains extra supersymmetry currents, enhancing $\NN=3$ to $\NN=4$ \cite{Aharony:2015oyb,Cordova:2016xhm}.}
Despite having no exactly marginal deformations, thus making them hard to study by the traditional field-theoretic approaches, various examples of non-trivial, pure, $\NN=3$ SCFTs have been constructed by now \cite{Garcia-Etxebarria:2015wns,Aharony:2016kai,Garcia-Etxebarria:2016erx} using  string-theoretic technology.
These theories were also recovered, and new ones obtained, by the systematic study of $\NN=2$ SCFTs with a one complex dimensional Coulomb branch in the work of \cite{Argyres:2015ffa, Argyres:2015gha, Argyres:2016xua, Argyres:2016xmc}.

Nevertheless, we still seem far from having a complete classification of $\NN=3$ SCFTs. One can hope that the situation is more tractable than the $\NN=2$ case, due to the extra supersymmetry, yet richer than $\NN=4$ where we might already have the complete classification.
Some of the known $\NN=3$ theories are obtained from $\NN=4$ SYM by gauging a discrete subgroup which, as pointed out in \cite{Aharony:2016kai,Argyres:2016yzz}, does not change the correlation functions nor the central charges of the theory, changing only the spectrum of local and non-local operators.
Among all non-trivial (\ie, that do not come from discrete gauging) pure $\NN=3$ theories known to date, the one
with the smallest central charge, and thus in a sense the simplest theory, has $a=c=\tfrac{15}{12}$.\footnote{In the notation of \cite{Aharony:2016kai} this corresponds to $N=1$ and $\ell=k=3$.}
One could wonder if this indeed corresponds to the ``minimal'' theory, or if there is a theory with lower central charge, perhaps not obtainable from S-fold constructions (and their generalizations). Thus we shall try to address these questions by field theoretic methods, and refrain from making any assumptions about the theories, apart from that it is a local and interacting $\NN=3$ SCFT.

\subsection{\texorpdfstring{Chiral algebra constraints on four-dimensional $\NN=3$ SCFTs}{Chiral algebra constraints on four-dimensional N=3 SCFTs}}

We take a bootstrap approach, bypassing the need for any perturbative description and making only use of the fact that any local $\NN=3$ SCFTs will have a stress tensor. The existence of the stress-tensor operator, together with all other operators that sit in the same $\NN=3$ superconformal multiplet, is the \emph{minimal} assumption one can make about local $\NN=3$ theories.\footnote{In the notation of \cite{Cordova:2016xhm,Cordova:2016emh} the stress tensor multiplet is denoted by $B_1 \bar{B}_1 (0,0)_{[1,1],0}^{2}$ and by $\hat{\BB}_{[1,1]}$ in \cite{Lemos:2016xke}.}
Therefore the constraints we obtain in this section are valid for \emph{any} $\NN=3$ SCFT and do not rely on any string-theoretic construction. We also do not assume any information about the Coulomb branch of the theory.
A downside of making the minimal set of assumptions about the theory is that we cannot impose it only has $\NN=3$ supersymmetry.
By simply considering the $\NN=3$ stress tensor four-point function we cannot distinguish between non-trivial $\NN=3$ SCFTs, and theories which are either $\NN=4$ theories or $\NN=3$ theories obtained from $\NN=4$ ones by gauging discrete symmetries.

To be able to rule out $\NN=4$ solutions one would have to impose that the multiplets containing the additional supercurrents, enhancing the symmetry from $\NN=3$ to $\NN=4$, are absent. However such multiplets are not exchanged in the most universal OPEs such as the stress tensor self-OPE.
This limitation can be overcome if one wants to construct the explicit chiral algebra of an $\NN=3$ SCFT, as done in \cite{Nishinaka:2016hbw,Lemos:2016xke}, but that requires making assumptions about the \emph{complete} list of generators of the chiral algebra, and thus is well suited to studying \emph{specific} known $\NN=3$ theories, but not to exploring the allowed space of $\NN=3$ SCFTs.\footnote{An attempt  to reach a compromise between the two options was explored in \cite{Lemos:2016xke} by constructing a candidate subalgebra of a large class of known $\NN=3$ SCFTs.}

\subsubsection*{The stress tensor multiplet}

Unsurprisingly, the operators in the stress-tensor multiplet of four-dimensional $\NN=3$ SCFTs give rise in cohomology to a two-dimensional $\NN=2$ stress-tensor multiplet. This corresponds to a long multiplet in two dimensions, $\TT(z,\t, \tb)= J(z) + G(z) \t + \bar{G}(z) \tb + \t \tb T(z)$, therefore requiring precisely the blocks computed in section \ref{sec:superblocks}.

The four-dimensional origin of each of the global conformal primaries in the superfield $\TT(z,\t, \tb)$ becomes more transparent if we view the $\NN=3$ theory as an $\NN=2$ one.
When viewed as an $\NN=2$ theory the $\U(3)_R$ R-symmetry group of $\NN=3$ theories decomposes as $\U(1)_F \times \U(2)_R$, where the first factor is the R-symmetry of the  $\NN=2$ superconformal algebra, while the second factor corresponds to a global symmetry from the $\NN=2$ point of view, \ie, it commutes with the $\NN=2$ superconformal algebra.
Decomposing the $\NN=3$ stress tensor multiplet in $\NN=2$ representations one finds
\begin{itemize}
\item the  $\U(1)_F$ flavor current multiplet ($\hat{\BB}_1$ in the notation of \cite{Dolan:2002zh}),
\item the stress tensor multiplet ($\hat{\CC}_{0,(0,0)}$), and
\item two supercurrent multiplets, containing extra currents enhancing $\NN=2$ to $\NN=3$ ($\DD_{1/2,(0,0)}$ and $\bar{\DD}_{1/2,(0,0)}$).
\end{itemize}
As described above the $\U(1)_F$ flavor symmetry gives rise in cohomology to a $\U(1)$ AKM current algebra, whose generator is given precisely by the dimension one superconformal primary of $\TT(z,\t,\tb)$: $J(z)$. The $\NN=2$ stress tensor multiplet gives rise to the two-dimensional stress tensor $T(z)$, while the extra supercurrents furnish $G(z)$ and $\bar{G}(z)$ \cite{Beem:2013sza}.
All of these two-dimensional global conformal primaries are related by the action of four of the extra supercharges (two Poincar\'{e} supercharges and their conjugates) which appear in the $\NN=3$ in addition to those of the $\NN=2$ subalgebra, and which commute with $\qq$.

\subsubsection*{Four-dimensional OPE coefficients from the chiral algebra}

We decomposed the four-point function of the two-dimensional $\NN=2$ stress-tensor multiplet $\TT(z,\t,\tb)$ in superblocks in section \ref{sec:STdecomp}.
Interpreting this decomposition in the context of the two-dimensional chiral algebra, each two-dimensional global superconformal primary operator arises as the representative of a
four-dimensional superconformal multiplet. Thus, the two-dimensional OPE coefficients obtained in this way amount to the computation of an infinite number of four-dimensional OPE coefficients.
Furthermore, even though the two-dimensional chiral algebra is not unitary, implying the sign of the two-dimensional OPE coefficients have a priori no constraint. By re-interpreting these OPE coefficients in a four-dimensional language we can impose unitarity of the four-dimensional theory and constrain their sign. This constrains which chiral algebras can arise from four-dimensional $\NN=3$ SCFTs.

Although the selection rules for the four-dimensional OPE of two $\NN=3$ stress-tensor multiplets remain elusive, and obtaining them is a project in itself, we can leverage knowledge of selection rules for $\NN=2$ SCFTs to interpret the computed two-dimensional OPE coefficients in terms of four-dimensional ones.
The superconformal primary of the two-dimensional stress tensor multiplet is the aforementioned AKM current. In four-dimensional language it arises from an $\NN=2$ $\hat{\BB}_1$ multiplet, whose OPE selection rules were obtained in \cite{Arutyunov:2001qw}
\be
\hat{\BB}_1 \times \hat{\BB}_1 = \mathcal{I} + \hat{\BB}_1 + \hat{\BB}_2 + \sum_{\ell=0}^\infty \hat{\CC}_{0,\ell} + \sum_{\ell=0}^\infty \hat{\CC}_{1,\ell}\,.
\label{eq:BhatOPE}
\ee
Here we only listed multiplets containing Schur operators, and thus relevant for our computation.
Of these multiplets the $\hat{\CC}_{0,\ell}$ with spin $\ell >0$ contain conserved currents of spin larger than two, which are expected to be absent in interacting theories \cite{Maldacena:2011jn,Alba:2013yda}. As such we set their OPE coefficients to zero by hand, thereby restricting only to \emph{interacting} theories.
We point out even though we are interpreting these OPE coefficients in terms of four-dimensional $\NN=2$ representations, by decomposing the full correlation in two-dimensional superblocks, the four-dimensional $\NN=2$ multiplets were organized in $\NN=3$ representations. In other words, the superblock decomposition allows us to identify which two-dimensional multiplets are global superconformal primaries, and which are global superdescendants, thereby identifying which $\NN=3$ multiplet each $\NN=2$ multiplet belongs to.
Recall that the OPE coefficients $a_{h_{\ex}}$, $b_{h_{\ex}}$, $c_{h_{\ex},q_{\ex=\pm 1}}$
in eq.\ \eqref{eq:hatf_ans} correspond to a global superprimary, the $G_{-1/2} \overline{G}_{-1/2}$ descendant, and $G_{-1/2}/\overline{G}_{-1/2}$ descendants, respectively. Therefore it is straightforward to identify which $\NN=3$ multiplet is being exchanged by making use of the decomposition of $\NN=3$ in $\NN=2$ of \cite{Lemos:2016xke}. The relevant decompositions are
\begin{align}
\label{eq:N3toN2dec}
\begin{alignedat}{2}
&\hat{\CC}_{[1,1],(\tfrac{\ell}{2},\tfrac{\ell}{2})}\rightarrow\; \hat{\CC}_{1,(\tfrac{\ell}{2},\tfrac{\ell}{2})} \oplus \hat{\CC}_{1,(\tfrac{\ell+1}{2},\tfrac{\ell + 1}{2})} \,,\qquad
& &\hat{\BB}_{[1,1]}\rightarrow \;  \hat{\BB}_{1} \oplus \hat{\CC}_{0,(0,0)}\,,\\
&\hat{\CC}_{[2,0],(\tfrac{\ell}{2},\tfrac{\ell+1}{2})}\rightarrow \hat{\CC}_{1,(\tfrac{\ell+1}{2},\tfrac{\ell+1}{2})} \,,\qquad
& &\hat{\BB}_{[2,2]}\rightarrow  \hat{\BB}_{2} \oplus \hat{\CC}_{1,(0,0)}\,,
\end{alignedat}
\end{align}
where we followed the labeling of $\NN=3$ multiplets of \cite{Lemos:2016xke}, and restricted the decompositions to the types of Schur multiplets exchanged in eq.\ \eqref{eq:BhatOPE}

All in all, we obtain from the two-dimensional OPE coefficients in eq.\ \eqref{eq:STdecinsuper}, the following four-dimensional OPE coefficients
\begingroup
\allowdisplaybreaks[1]
\begin{align}
\label{eq:OPEB11}
&\lambda^2_{\hat{\BB}_{[1,1]}\;\mathrm{desc.}}
=- \frac{2}{c_{2d}}\,,\\
\label{eq:OPEB22}
&\lambda^2_{\hat{\BB}_{[2,2]}\;\mathrm{prim.}}
=2-\frac{2}{c_{2d}}\,,\\
\label{eq:OPEC11}
&\lambda^2_{\hat{\CC}_{[1,1],\ell}\;\mathrm{prim.}}
=\frac{3 \sqrt{\pi } 2^{-2 \ell -3} (\ell  (\ell +4)+1) \Gamma (\ell +3)}{c_{2d} (\ell +3) \Gamma \left(\ell +\frac{7}{2}\right)}+\frac{\sqrt{\pi } 2^{-2 \ell -5} (\ell +2) (\ell +4) (\ell +5) \Gamma (\ell +3)}{\Gamma \left(\ell +\frac{7}{2}\right)}\,, \quad \ell \; \mathrm{odd}\,,\\
\label{eq:OPEC11des}
&\lambda^2_{\hat{\CC}_{[1,1],\ell-1}\; \mathrm{desc.}} 
= \frac{3 \sqrt{\pi } 2^{-2 \ell -3} (\ell  (\ell +6)+6) \Gamma (\ell +2)}{c_{2d} (\ell +2) \Gamma \left(\ell +\frac{7}{2}\right)}+\frac{\sqrt{\pi } 2^{-2 \ell -5} \ell  (\ell +1) \Gamma (\ell +4)}{(\ell +2) \Gamma \left(\ell +\frac{7}{2}\right)}\,, \quad \ell \; \mathrm{odd}\,,\\
\label{eq:OPEC20}
&|\lambda_{\hat{\CC}_{[2,0],\left(\tfrac{\ell-1}{2},\tfrac{\ell}{2}\right)}\; \mathrm{desc.}}|^2 
=\frac{3 \sqrt{\pi } 2^{-2 \ell -3} (\ell +1) (\ell +4) \Gamma (\ell +2)}{c_{2d} (\ell +3) \Gamma \left(\ell +\frac{5}{2}\right)}+\frac{\sqrt{\pi } 2^{-2 \ell -5} (\ell +1) \Gamma (\ell +5)}{(\ell +3) \Gamma \left(\ell +\frac{5}{2}\right)}\,, \quad \ell \; \mathrm{odd}\,,
\end{align}
\endgroup
where the OPE coefficients in eqs.\ \eqref{eq:OPEB22} and \eqref{eq:OPEC11} correspond to the exchange of a two-dimensional global superprimary, eqs.\ \eqref{eq:OPEB11} and \eqref{eq:OPEC11des} to a $G_{-1/2} \overline{G}_{-1/2}$ descendant, and eq.\ \eqref{eq:OPEC20} to a $G_{-1/2}/\overline{G}_{-1/2}$.
We point out that we only know the OPE coefficient of the \emph{Schur} operator that is exchanged in the $\TT \TT$ OPE, which is not enough to obtain all four-dimensional OPE coefficients appearing in the full three-point function of two stress tensors and the multiplet in question.
The above, nonetheless, provides the subset of the selection rules of $\NN=3$ stress tensor multiplet OPE that is captured by the chiral algebra.

\subsubsection*{A new $\NN=3$ unitarity bound}

Unitarity requires all of the above coefficients to be positive, implying lower bounds on the value of $c_{4d}=-\tfrac{c_{2d}}{12}$, with the strongest one coming from the OPE coefficient $\lambda^2_{\hat{\CC}_{[1,1],\ell=0}\; \mathrm{desc.}}$ ($=b_{h_{\ex}=3}$), \ie, the exchange of a dimension four descendant (a $\hat{\CC}_{1,\ell=1}$ multiplet) of an uncharged global superprimary of dimension three (a $\hat{\CC}_{1,\ell=0}$ multiplet).
This yields the following unitarity bound
\be
c_{4d} \geqslant \frac{13}{24}\,,
\label{eq:cbound}
\ee
valid for any local \emph{interacting} $\NN=3$ SCFT.
Unlike the previous unitarity bounds obtained from chiral algebra correlators \cite{Beem:2013qxa,Beem:2013sza,Liendo:2015ofa,Lemos:2015orc,Beem:2016wfs}
the inequality \eqref{eq:cbound} is not saturated by any known theory, and in fact we will argue that the bound is a strict inequality. Similar bounds, relying only on the existence of a stress tensor, and the absence of higher spin currents, for theories with $\NN=2$ and $\NN=4$ supersymmetry ($c_{4d} \geqslant \tfrac{11}{30}$ and $c_{4d} \geqslant \tfrac{3}{4}$ respectively \cite{Liendo:2015ofa,Beem:2013qxa,Beem:2016wfs}), are saturated by the interacting theories with lowest central charge known in each case: the simplest Argyres-Douglas point \cite{Argyres:1995jj,Argyres:1995xn} for the former, and by $\NN=4$ super Yang-Mills (SYM) with gauge group $\SU(2)$ for the latter, but we claim this cannot happen for \eqref{eq:cbound}. Moreover there is no known SCFT whose central charge is close to saturating it. In particular, there is no known theory with central charge in between this value and that of $\NN=4$ SYM with gauge group $\SU(2)$. These values are below those values that were seen in the systematic classification of theories with a one-dimensional Coulomb branch of \cite{Argyres:2015ffa, Argyres:2015gha, Argyres:2016xua, Argyres:2016xmc}.
Moreover, making use of eq.~(5.1) of \cite{Argyres:2016xmc} one can obtain, under certain assumptions, including that the Coulomb branch is freely generated, that for a rank one $\NN=3$ SCFT $c_{4d} \geqslant \tfrac{3}{4}$.\footnote{We thank Mario Martone for discussions on this point.} A theory close to saturating \eqref{eq:cbound} would then seem to be an interacting rank zero SCFT (\ie, with no Coulomb branch) or with non-freely generated Coulomb branches.\footnote{In \cite{Font:2016odl} six-dimensional theories were found that could have rank zero, although this was not the only possibility there, we thank I.~Garc\'{i}a-Extebarria for bringing this reference to our attention.}

\subsubsection*{Reconstructing $4d$ operators from the chiral algebra}

We should emphasize that it is still not clear what is the full set of conditions a two-dimensional chiral algebra must satisfy
such that it arises from a consistent four-dimensional SCFT. In the case at hand, however, we will give an argument as to why
an $\NN=2$ chiral algebra with $c_{2d}=-13/2$ cannot correspond to an \emph{interacting} four dimensional
$\NN=3$ SCFT.

Let us suppose that there exists an \emph{interacting} four-dimensional SCFT for some given value of $c_{4d}$. Then we can construct
in the chiral algebra the operators that are exchanged in the $\mathcal{T} \mathcal{T}$ OPE. In our discussion here we will focus on
uncharged dimension three global superprimaries, since the bound \eqref{eq:cbound} arises from the exchange of a superdescendant of
such an operator. From four-dimensional selection rules we know the global superprimaries of the operators being exchanged have to
belong in $\hat{\CC}_{[1,1],\ell=0}$ or  $\hat{\CC}_{[0,0],\ell=1}$ representations, and we impose the latter to be absent to focus
on interacting theories. When passing to the cohomology of \cite{Beem:2013sza}, Schur operators from different $4d$ multiplets can
give rise to global supermultiplets of the 2-dimensional $\NN=2$ algebra that look identical. In particular they may contain
two-dimensional superprimaries of the same weight and $U(1)_F$ (and also $U(1)_r$) charges. One such example is given by the
$\NN=2$ multiplets $\hat{\CC}_{0,\ell}$ and $\hat{\CC}_{1,\ell-1}$ in a four-dimensional theory. 

For the arguments we outline below it will be crucial to distinguish between $4d$ multiplets that give rise to identical
superconformal multiplets in cohomology. The ambiguities that can appear were discussed in \cite{Lemos:2016xke}, and for 
theories with a single chiral algebra generator a conjectural prescription on how to lift them was put forward in 
\cite{Song:2016yfd}. Since such prescription does not apply to the case at hand we simply exploit that cohomology inherits a bit more 
structure from the reduction process than the spectrum of charges and weights. Namely, it also induces an indefinite 
quadratic form. Orthogonal $4d$ multiplets remain so in cohomology, but their superprimaries may give rise to states 
of negative norm. In this way, it may be possible to distinguish between two multiplets with identical spectra of 
weights and charges. This is the case for the $\NN=2$ multiplets relevant here (see \eqref{eq:N3toN2dec}) $\hat{\CC}_{1,0}$
and $\hat{\CC}_{0,1}$ 
\cite{Beem:2013sza} which indeed reduce to identical superprimaries, but with norms of opposite signs.
When we reduce the stress tensor operator product expansion of the four-dimensional theory we obtain the superdescendant of a $2d$ uncharged superconformal primary of dimension $h=3$ which has negative norm with respect to the induced 
quadratic form. 

Let us now look at the two-dimensional side of the story. For central charges around the value $c_{2d}=-\tfrac{13}{2}$, the 
subspace of uncharged dimension $h=3$ superprimaries is 2-dimensional and its quadratic form is indefinite, \ie, it 
possesses one positive and one negative eigenvalue. Let us stress that both eigenvalues are non-zero. Given any 
choice of an orthonormal basis $\OO_1 $ and $\OO_2$ of this space, we can reach any other choice
by an SO$(1,1)$ transformation. Let us denote the unique parameter of SO$(1,1)$ by $b$ and the corresponding basis vectors by $\OO_1(b)$ and
$\OO_2(b)$. Without loss of generality we can assume that the vectors $\OO_1(b)$ are those with negative norm while the norm of $\OO_2(b)$
is positive. According to our previous discussion, we must show the operator product expansion of the stress tensor in the 2-dimensional
theory contains the global superdescendant of $\OO_1(b)$ and is orthogonal to the global superdescendant of $\OO_2(b)$ for some choice of the parameter $b$. This is indeed possible for all values
of the central charge $c_{2d} > - \tfrac{13}{2}$. In fact, one can show that the 3-point functions
$$ \langle \mathcal{T}(w_1) \mathcal{T}(w_2) G \overline{G} \OO_2(b;z)\rangle = 0 \quad \mbox{ for some } \quad b = b(c_{2d}) \ , $$
where $G \overline{G}$ means we are referring to the dimension four superdescendant of $\OO_2(b;z)$.
The negative norm field $\OO_1$, whose descendant appears in the operator product expansion, is given by
\be
\OO_1(b(c_{2d});z) = 2 \left(6(\bar{G} G)(z) - (J T)(z)+J(z)''\right) -\frac{3}{2} T(z)'\,.
\ee
Here $(AB)(z)$ means the normal-ordered product of $A(z)$ and $B(z)$. The dimension four superdescendant of $\OO_1(b,z)$ corresponds to
the operator exchanged in the $\TT \TT$ OPE. When $c_{2d}$ approaches the value $c_{2d}=-\tfrac{13}{2}$, however, the boost parameter $b$ tends to infinity and
the field $\OO_1(b(c_{2d}=-\tfrac{13}{2});z)$ has vanishing 2-point function. This means that the stress tensor OPE at $c_{2d}=-\tfrac{13}{2}$ is inconsistent with the
cohomological reduction from a four-dimensional interacting $\NN=3$ theory with central charge $c_{4d} = \tfrac{13}{24}$. Hence we conclude that such
a theory cannot exist. Let us stress, though, that our argument relies on one additional assumption, namely that the quadratic form in
cohomology coincides with the usual Shapovalov form in the vacuum sector of the $\NN=2$ Virasoro algebra. This is not guaranteed, much   
as it is not guaranteed that the global $\NN=2$ superconformal symmetry that acts on cohomology is enhanced to a super Virasoro symmetry. 
On the other hand, such an enhancement is seen in many explicit examples and it seems natural to expect that it extends to the relevant 
quadratic form. 

An immediate question that arises is whether our arguments could be refined to obtain a bound stronger than \eqref{eq:cbound}. In particular
there is no known $\NN=3$ theory whose chiral algebra is generated only by the stress tensor multiplet $\TT$ (the $\NN=2$ and $\NN=4$
SCFTs with smallest central charge have as chiral algebras the (super) Virasoro vacuum module), could such a theory exist? A necessary
condition for this to happen would be the existence of a null state in the chiral algebra involving a power of the stress tensor as
discussed in \cite{ChrisLeonardounpublished}.

\section{Conclusions}
\label{sec:conclusions}

The superconformal bootstrap program has been very successful in recent years, allowing for non-perturbative results to be obtained in theories hard to access by other means.  To  make progress in achieving its two ambitious goals, of charting out the theory space, and solving specific theories, one must start considering less supersymmetric multiplets.
In this work we initiated, in two dimensions, the bootstrap of long multiplets, using the whole superfield as the external operator.
While long multiplets have been considered in the past, from the point of view of kinematics \cite{Fortin:2011nq,Berkooz:2014yda,Khandker:2014mpa,Li:2014gpa,Bobev:2015jxa,Li:2016chh}, and recently through a numerical analysis of dynamical information \cite{Li:2017ddj},
all previous work has been restricted to considering only the superconformal primary
of long multiplets.
Unlike the case of external chiral operators (or BPS operators in general) where the four-point function depends only on the supersymmetrization of the regular bosonic conformal and R-symmetry invariants, for more general external fields one starts finding nilpotent superconformal invariants. This implies that information is lost by restricting the four-point function to the superconformal primary, \ie, setting all fermionic coordinates to zero.
Even in cases such as the one considered in section \ref{sec:numerics}, where  Bose symmetry fixes the correlation function involving external descendants from that of external primaries, the crossing symmetry constraints for correlators with external superdescendants were nontrivial. Upon setting all fermionic coordinates to zero the superblocks restrict to bosonic conformal blocks, and the crossing equations are simply those of a non-supersymmetric theory. Supersymmetry manifests itself in the constraints appearing when one considers also external superdescendents.

Although we treated the two-dimensional $\NN=1$ case only as a warm-up example, making manifest some of the features important to our discussion, we expect that also in the case of the $\NN=(1,1)$ bootstrap, non-trivial constraints arise from considering the full four-point function. As was pointed out in \cite{Bobev:2015jxa} for a two-dimensional $\NN=(1,1)$ SCFT, if we restrict the external operators to be the superconformal primaries, \ie, setting all fermionic coordinates to zero, the superconformal blocks reduce to a sum of bosonic blocks.\footnote{Similarly for the case of three-dimensional $\NN=1$, which has the same number of supercharges, the superblock turns into a regular bosonic block after setting all fermionic variables to zero  \cite{Bashkirov:2013vya}.} Once again the non-trivial constraints should come from considering the correlation functions of external superdescendants.

In two-dimensions, the blocks obtained in this work are restricted to the OPE channel between opposite charged
operators for brevity, but it would be straightforward to obtain results in the OPE channel between operators
of different charge. The charged sector, and hence the full set of superblocks for any value of the external
charges, is important of one wants to distinguish the (2,2) minimal models for the (2,0) heterotic theories
\eqref{Zk} of Gadde and Putrov. Of course, studying the space of (2,2) theories is of independent interest.
The numerical bootstrap approach to $\NN=(2,2)$ case can be easily addressed simply by patching together the
holomorphic and anti-holomorphic superblocks of section \ref{sec:superblocks}, extending the work done in
\cite{Bobev:2015vsa} from chiral operators to long ones.

Finally, one clear future direction would be to extend these results to higher dimensions, following what was done in \cite{Bobev:2015jxa} for correlators involving chiral operators. In particular they defined the superconformal algebra with four supercharges in an arbitrary number of dimensions, allowing to write the Casimir operator in $2 \leqslant d \leqslant 4$. Recall that this corresponds to theories with $\NN=(2,2)$ in two dimensions and $\NN=1$ in four dimensions. By solving the Casimir equation in arbitrary $d$ one gets, in one blow, the superblocks involving chiral fields for all these theories. Our approach in this paper provides the case of the two-dimensional $\NN=(2,2)$ long blocks, and the structure of the four-point function, \ie, the number of superconformal invariants will be the same in higher dimensions. Therefore one could write both the quadratic and cubic Casimirs in arbitrary dimensions and proceed along the lines of section \ref{sec:superblocks}.
One technical difficulty we can foresee is the need to use spinning blocks, as even if one consider a scalar superconformal primary, among its superdescendants operators with spin will appear. Moreover, solving the Casimir equation is much easier if one can give an Ansatz for the superblock in terms of a sum of bosonic blocks. Such a procedure requires constructing conformal primaries out of superdescendants which can get cumbersome. Alternatively, it would be of interest to extend the approach
proposed in \cite{Schomerus:2016epl} to the case of superconformal groups. Quite generally, it leads to a reformulation of
conformal Casimir equations as eigenvalue equations for certain Calogero-Sutherland Hamiltonians, in agreement with
\cite{Isachenkov:2016gim}. As was shown at the example of three-dimensional fermionic seed blocks in \cite{Schomerus:2016epl},
the reformulation in terms of Calogero-Sutherland models is very universal and in particular works for spinning blocks
as well as for scalars. Hence, one would expect that a universal set of Casimir equations for long multiplets of
superconformal groups can be derived in any dimension. Moreover, by exploiting the integrability
of Calogero-Sutherland Hamiltonians it should be possible to develop a systematic solution theory \cite{Isachenkov:2016gim,Isachenkov:2017new}, without the need for an Ansatz that decomposes
superblocks in terms of bosonic ones.

Another clear future direction corresponds to obtaining the $\NN=1$ stress-tensor multiplet superblocks, which despite being an essential multiplet to consider in any bootstrap studies, remain unknown. In this case however the superconformal primary has spin one. It could happen that the extra conditions arising from conservation make the Casimir differential equations in this case simpler to solve, otherwise it could simply be obtained from imposing conservation on the generic long blocks.

\bigskip

In a different direction, the holomorphic long blocks we computed, plus the same blocks relaxing the charge conditions we took for simplicity, together with the blocks involving external chiral operators of \cite{Fitzpatrick:2014oza,Bobev:2015jxa} are all the blocks that are required for the study of chiral algebras \cite{Beem:2013sza} associated to $\NN=3$ SCFTs.
These blocks allowed us to obtain an infinite number of four-dimensional (sums of squared) OPE coefficients of $\NN=3$ theories, in terms of a single parameter, the central charge of the four-dimensional theory. These numbers correspond to the coupling between the Schur operators in the four-dimensional stress-tensor multiplet, and the Schur operators that appear in its self-OPE. They are universal, in the sense that no assumptions about specific $\NN=3$ theories were made, apart from the demand that the theory be interacting, and are a necessary ingredient in the superconformal bootstrap program of $\NN=3$ stress tensors.

Requiring unitarity of the four-dimensional theory provided a new analytic unitarity bound
\be
c_{4d} \geqslant \frac{13}{24}\,,
\label{eq:cboundconc}
\ee
valid for any interacting theory. Unlike similar bounds for $\NN=2$ and $\NN=4$ SCFTs, we have argued this bound \emph{cannot} be saturated by any interacting unitary SCFT. 
Our arguments have provided a first non-trivial example of a chiral algebra that 
cannot appear as cohomology of a four-dimensional SCFT.
Namely they provided an example of what can go wrong when we try to interpret a given chiral algebra as arising from a four-dimensional SCFT. Since there are also no known theories close to saturating \eqref{eq:cboundconc} one might wonder if they could be ruled out by reasonings similar to the one used here, and whether its possible to obtain a stronger bound saturated by a physical $4d$ SCFT. We leave this question for future work, as it would require going deeper in the bigger question of what are the requirements for a two-dimensional chiral algebra to correspond to fully consistent four-dimensional SCFT. Similar reasoning might also help improve the bounds obtained in \cite{Beem:2013sza,Lemos:2015orc}.
Adding extra assumptions about specific theories by considering mixed systems of correlators, such as including chiral operators (arising from four-dimensional half-BPS multiplets) could provide new constraints on the space of theories, although one starts getting ambiguities in the four-dimensional interpretation of two-dimensional multiplets, as discussed in \cite{Lemos:2016xke} for the simplest half-BPS correlator.

Finally, the blocks we have computed are a piece of the full four-dimensional superblocks of (non-chiral) Schur operators, obtained by performing the chiral algebra twist on the full blocks. An essential superblock for the $\NN=3$ superconformal bootstrap program  corresponds to
having stress-tensor multiplets as the external state. Although these blocks are still unknown, our analysis captures the chiral algebra subsector of these blocks, and in particular the statement that information is lost by setting all fermionic variables to zero (\ie, considering the correlation function of superconformal primaries) remains true for the whole system.

\medskip
\acknowledgments

We have greatly benefited from discussions with
P.~Argyres,
C.~Behan,
I.~Garc\'{i}a-Etxebarria,
P.~Liendo,
M.~Martone,
L.~Rastelli,
and
B.~van~Rees.
The research leading to these results has received funding from the People Programme (Marie Curie Actions) of the European Union’s Seventh Framework Programme FP7/2007-2013/ under REA Grant Agreement No 317089 (GATIS).
The authors thank the Galileo Galilei Institute for Theoretical Physics for hospitality and the INFN for partial support during the completion of this work during the workshop ``Conformal field theories and renormalization group flows in dimensions $d>2$''.


\appendix
\section{Casimir and crossing equations}
\label{app:blocks}

This appendix collects some lengthy equations used to obtain the $\NN=2$ superconformal blocks in section \ref{sec:superblocks} and the crossing equations for $\NN=(2,0)$ SCFTs in section \ref{sec:numerics}.

\subsection{Casimir equations}
\label{app:quadraticCasimireqs}

\subsubsection*{Quadratic Casimir differential equation}

The application of the 
quadratic Casimir (obtained from eq.\ \eqref{eq:Cas2}) to the four-point function, through the differential action \eqref{difform}  of the generators yields a system of six coupled differential equations for the six functions $f_i(z)$ in eq. \eqref{Fexp}.

After some rearrangements we find that two of the six functions are completely determined in terms of the function $f_0$
\begin{align}
\begin{split}
f_1(z) &= \frac{z^3 f_0''(z)-z^2 f_0''(z)+z^2 f_0'(z)+\CasEig_2 f_0(z)}{z}\,,\\
f_5(z)&=\frac{z^2 \left((z-1) \left((2 \CasEig_2+2 z-1) f_0''(z)+2 z f_1''(z)\right)+(2 \CasEig_2+2 z-1) f_0'(z)\right.}{z^2}  \\
&+\frac{\left.(6 z-4) f_1'(z)\right)+2 \CasEig_2 f_0(z) (\CasEig_2+z-1)}{z^2}\,,
\end{split}
\end{align}
and that the differential equations involving $f_0$ is totally decoupled and can be written in terms of (minus) the usual bosonic Casimir
\be
\CC_2=z^2 \left((z-1) \frac{\partial ^2 f(z)}{\partial z^2}+\frac{\partial f(z)}{\partial z}\right)\,,
\ee
as
\be
2 \DD(f_0) (\CasEig_2+4 z-2)+\frac{\partial ^2\left(2 \DD(f_0) (z-1) z^2\right)}{\partial z^2}+\frac{\partial (-2 \DD(f_0) z (5 z-4))}{\partial z^1}=0\,,
\ee
where
\be
\DD(f_0) = 2 \CasEig_2 \CC_2(f_0(z))+(\CasEig_2-1) \CasEig_2 f_0(z)+\CC_2(\CC_2(f_0(z)))\,.
\ee
The other three functions are determined by the following equations
\begin{align}
\begin{split}
&f_2(z)+zf_2'(z)+z^2 \left(-f_3''(z)\right)-2 z f_3'(z)-\frac{\CasEig_2 f_3(z)}{z-1} =0\,,\\
&f_2(z)+zf_2'(z)+(z-1) z^2 f_4''(z)+2 (2 z-1) z f_4'(z)+f_4(z) (\CasEig_2+2 z) =0\,,  \\
&-2 (\CasEig_2-1) f_2(z)-z f_3'(z)+2 (z-1)^2 z^3 f_4^{(3)}(z)+8 (z-1) (2 z-1) z^2 f_4''(z)  \\
&+ z \left(2 \CasEig_2 z-2 \CasEig_2+28 z^2-27 z+3\right) f_4'(z)+z f_4(z) (2 \CasEig_2+8 z-3)=0 \,. \\
\end{split}
\end{align}
Recall that the eigenvalue of the quadratic Casimir is $\CasEig_2=h_{ex}^2 - \tfrac{q_{ex}^2}{4}$, where $h_{ex}$ and $q_{ex}$ are the charges of the superconformal primary of the supermultiplet being exchanged.
This system is rather cumbersome to solve, and thus to solve it in section \ref{sec:longblockssol} we change ``basis'' from the functions  $f_i(z)$ defined in eq.\ \eqref{Fexp}, to functions $\hat{f}_i$ (defined in eq.\ \eqref{eq:ftofhat}) where one can more easily give an Ansatz in terms of a sum of bosonic blocks \eqref{eq:hatf_ans}.
The solution for the exchange of uncharged supermultiplets is collected in eqs.\ \eqref{eq:atypesol} and \eqref{eq:btypesol}, according to whether a superconformal 
primary or descendant is exchanged.

\subsubsection*{Cubic and quadratic Casimir equations for the charged exchange}

As clear from the quadratic Casimir eigenvalue the equations in appendix \ref{app:quadraticCasimireqs} do not distinguish between the exchange of a superconformal multiplet with positive or negative charge, and thus we need also to consider the cubic Casimir \eqref{Cas3}.
Considering these two equations suffices to fix all parameters in the Ansatz \eqref{eq:hatf_ans}, giving the solution in eq.\ \eqref{eq:ctypesol}, and the quartic Casimir gives no new information.
However some of the equations arising from the quartic Casimir appear in a simpler form, and using them we can easily simplify the system of Casimir equations, solving for all $\hat{f}_i(z)$ in terms of $\hat{f}_0(z)$,\footnote{Note that we always assume that the external fields are not chiral.}
\begingroup
\allowdisplaybreaks[1]
\begin{align}
\label{quarticeq}
\begin{split}
\hat{f}_1(z)&= \frac{z^2 (2 h +q_1) \left(\hat{f}_0'(z)+(z-1) \hat{f}_0''(z)\right)}{\CasEig_2}\,,\\
\hat{f}_2(z)&= -\frac{z^2 (2 h -q_1) \left(\hat{f}_0'(z)+(z-1) \hat{f}_0''(z)\right)}{\CasEig_2}\,, \\
\hat{f}_3(z)&= \frac{(z-1) z (q_1+2 h  q_{ex}) \hat{f}_0'(z)}{2 h }\,, \\
\hat{f}_4(z)&= \frac{z (2 h  q_{ex}-q_1) \hat{f}_0'(z)}{2 h }\,, \\
\hat{f}_5(z)&= -\frac{z^2 \left(\CasEig_2 \left(4 h ^2+q_1^2\right)-8 h ^2 \left(-4 h ^2+q_1^2+\CasEig_4\right)\right) \left(\hat{f}_0'(z)+(z-1) \hat{f}_0''(z)\right)}{4 h ^2 \CasEig_2}  \\
&+\frac{2 h  (2 h -1) \CasEig_2 \left(4 h ^2-q_1^2\right) \hat{f}_0(z)}{4 h ^2 \CasEig_2}\,,
\end{split}
\end{align}
\endgroup
where $\CasEig_4= q_{ex}^2 \CasEig_2$, and find a differential equation for $\hat{f}_0(z)$ only
\be
\CasEig_2 \hat{f}_0(z)+z^2 \left(\hat{f}_0'(z)+(z-1) \hat{f}_0''(z)\right)=0\,.
\ee
We recognize  this equation as the bosonic Casimir equation with eigenvalue $h(h-1)= \CasEig_2$, whose solution, for $q_{ex}= \pm 1$, is simply given by the  $\sl(2)$ bosonic block with holomorphic dimension $h_{ex}+\tfrac{1}{2}$.
Inserting this solution into eq.\  \eqref{quarticeq} gives immediately the result for the functions $\hat{f}_i$ given in eq.\ \eqref{eq:ctypesol}, and all other equations arising from the system of Casimirs are satisfied.

\subsection{\texorpdfstring{$\NN=(2,0)$}{N=(2,0)} crossing equations}

Here we collect the Taylor expansion of the crossing equations \eqref{eq:crossing} in the nilpotent invariants ($I_{i \neq 0}$),
\begin{align}
\begin{split}
\label{eq:crossing_expanded}
&(\zb-1)^{2\hb} \left(g_0(I_0,\zb)+ I_1 g_1(I_0,\zb)+I_2 g_2(I_0,\zb)+I_3 g_3(I_0,\zb)+I_4 g_4(I_0,\zb) -I_3 I_4 (1-I_0)g_5(I_0,\zb)\right) =  \\
& I_0^{2h} \zb^{2\hb} \Bigg( g_0(I_0^{-1},1-\zb) + \frac{I_1}{I_0}\bigg(2h g_0(I_0^{-1},1-\zb)+\big(1-\frac{1}{I_0}\big)g_0'(I_0^{-1},1-\zb)-g_1(I_0^{-1},1-\zb)\bigg)\\
&- \frac{I_2}{I_0} \bigg( g_2(I_0^{-1},1-\zb)+ g_3(I_0^{-1},1-\zb)\bigg) +\frac{I_3}{I_0} g_3(I_0^{-1},1-\zb)+ \frac{I_4 (1-I_0)}{I_0} \bigg(2g_2(I_0^{-1},1-\zb)\\
& +g_3(I_0^{-1},1-\zb)+g_4(I_0^{-1},1-\zb)  \bigg) +\frac{I_3I_4(1-I_0)}{I_0^2} \bigg( 2h(2 h-1)g_0(I_0^{-1},1-\zb)+\big(1-\frac{2}{I_0}+\\
&\frac{1}{I_0^2}\big)g_0''(I_0^{-1},1-\zb) -2(2 h-1)g_1(I_0^{-1},1-\zb)-2\big(1-\frac{1}{I_0}\big)g_1'(I_0^{-1},1-\zb)
+g_5(I_0^{-1},1-\zb)\bigg)\Bigg )\,,
\end{split}
\end{align}
with the coefficient of each invariant giving rise to a crossing equation, as discussed in section \ref{sec:crossing}, ultimately culminating in the crossing equation \eqref{eq:final_crossing}. 

\bibliography{./auxi/biblio}
\bibliographystyle{./auxi/JHEP}

\end{document}